\documentclass[12pt]{article}
\pdfoutput=1
\usepackage{epsf}

\usepackage{epsfig,graphics}
\usepackage {graphicx}
\usepackage {epsfig}
\usepackage {subfigure}
\usepackage {tabularx} 
\usepackage{rotate}	
\usepackage{slashed}

\usepackage{amsmath}
\usepackage{amsfonts}
\usepackage{amssymb}
\usepackage{dsfont}
\usepackage{graphicx}
\usepackage{cite}
\usepackage{url}
\usepackage{hyperref}
\hypersetup{colorlinks=false,linkcolor=black}
\usepackage{pgfplots}
\pgfplotsset{compat=newest}

\definecolor{myblue}{cmyk}{0.65, 0.37, 0.0, 0.19}

\newcommand{\bmat}{\left(\begin{array}}
\newcommand{\emat}{\end{array}\right)}

\def\yzero{\smash{\hbox{$y\kern-4pt\raise1pt\hbox{${}^\circ$}$}}}

\def\beq{\begin{equation}}
\def\eeq{\end{equation}}
\def\beqa{\begin{eqnarray}}
\def\eeqa{\end{eqnarray}}

\def\-{\hphantom{-}}

\def\s2{\frac{1}{\sqrt2}}

\def\beq{\begin{equation}}
\def\eeq{\end{equation}}
\def\beqa{\begin{eqnarray}}
\def\eeqa{\end{eqnarray}}

\def\IF{\relax{\rm I\kern-.18em F}}
\def\II{\relax{\rm I\kern-.18em I}}
\def\IP{\relax{\rm I\kern-.18em P}}
\def\IC{\relax\hbox{\kern.25em$\inbar\kern-.3em{\rm C}$}}
\def\IR{\relax{\rm I\kern-.18em R}}

\def\Dsl{\,\raise.15ex\hbox{/}\mkern-13.5mu D} 
\def\IZ{Z\kern-.4em  Z}

\catcode`\@=11   
\newdimen\@rotdimen
\newbox\@rotbox  

\def\@vspec#1{\special{ps:#1}}
\def\@rotstart#1{\@vspec{gsave currentpoint currentpoint translate
   #1 neg exch neg exch translate}}
\def\@rotfinish{\@vspec{currentpoint grestore moveto}}
\def\@rotr#1{\@rotdimen=\ht#1\advance\@rotdimen by\dp#1%
   \hbox to\@rotdimen{\hskip\ht#1\vbox to\wd#1{\@rotstart{90 rotate}%
   \box#1\vss}\hss}\@rotfinish}
\def\@rotl#1{\@rotdimen=\ht#1\advance\@rotdimen by\dp#1%
   \hbox to\@rotdimen{\vbox to\wd#1{\vskip\wd#1\@rotstart{270 rotate}%
   \box#1\vss}\hss}\@rotfinish}%
\def\@rotu#1{\@rotdimen=\ht#1\advance\@rotdimen by\dp#1%
   \hbox to\wd#1{\hskip\wd#1\vbox to\@rotdimen{\vskip\@rotdimen
   \@rotstart{-1 dup scale}\box#1\vss}\hss}\@rotfinish}%
\def\@rotf#1{\hbox to\wd#1{\hskip\wd#1\@rotstart{-1 1 scale}%
   \box#1\hss}\@rotfinish}%
\def\rotate{\@ifnextchar[{\@rotate}{\@rotate[l]}}
\def\@rotate[#1]#2{\setbox\@rotbox=\hbox{#2}\@nameuse{@rot#1}\@rotbox}

\catcode`\@=12

\topmargin
-1.5cm
\textwidth
15.5cm
\textheight
23.5cm
\oddsidemargin
0.7cm
\evensidemargin
0.7cm

\begin{document}

\makeatletter
\@addtoreset{equation}{section}
\makeatother
\renewcommand{\theequation}{\thesection.\arabic{equation}}
\pagestyle{empty}
\vspace{-0.2cm}
\rightline{IFT-UAM/CSIC-15-51}
\rightline{BONN-TH-2017-05, MPP-2017-123}
\vspace{-0.2cm}
\begin{center}


\LARGE{Constraining Neutrino Masses,\\the Cosmological Constant
and BSM Physics from the Weak Gravity Conjecture} 
\\[2mm]
  \large{    Luis E. Ib\'a\~nez$^1$,
    V\'ictor Mart\'in-Lozano$^2$,
    Irene Valenzuela$^{3,4}$  \\[3mm]}
\small{
  $^1$Departamento de F\'{\i}sica Te\'orica
and Instituto de F\'{\i}sica Te\'orica UAM/CSIC,\\[-0.3em]
Universidad Aut\'onoma de Madrid,
Cantoblanco, 28049 Madrid, Spain \\[0pt]
      $^2$Bethe Center for Theoretical Physics \& Physikalisches Institut der Universit\"{a}t Bonn,\\ Nu{\ss}allee 12, 53115, Bonn, Germany\\[0pt]
      $^3$Max Planck Institute for Physics,
     F\"ohringer Ring 6, 80805 Munich, Germany \\[0pt]
      $^4$Institute for Theoretical Physics and Center for Extreme Matter and Emergent Phenomena\\
      Utrecht University, Leuvenland 4, 3584 CE Utrecht, The Netherlands
\\[2mm]
}
\small{\bf Abstract} \\ [3mm]
\end{center}
\begin{center}
\begin{minipage}[h]{15.22cm}
It is known that there are AdS vacua obtained from compactifying the 
SM to 2 or 3 dimensions. The existence of such vacua depends on the value of neutrino masses  through the Casimir effect.
Using the Weak Gravity Conjecture, it has been recently argued by Ooguri and Vafa that such vacua are incompatible with
the SM embedding into a consistent theory of quantum gravity. 
We study the limits obtained for both the cosmological constant $\Lambda_4$ and neutrino masses from  the absence of  such dangerous  3D and 2D
SM  AdS vacua. One interesting implication  is that  $\Lambda_4$ is bounded to be larger than a scale of order $m_\nu^4$, as observed experimentally. 
Interestingly, this  is the first argument implying a non-vanishing $\Lambda_4$  only on the basis of particle physics, with no cosmological input.
Conversely, the observed  $\Lambda_4$ implies strong constraints 
on neutrino masses in the SM and also for some BSM extensions including  extra Weyl or Dirac 
spinors, gravitinos and axions.
The upper bounds obtained for neutrino masses imply (for fixed neutrino Yukawa and $\Lambda_4$) the existence of upper bounds on the EW scale.
In  the case of massive 
Majorana neutrinos with a see-saw mechanism associated to a large scale  $M\simeq 10^{10-14}$ GeV and $Y_{\nu_1}\simeq 10^{-3}$, one obtains 
that the EW scale cannot exceed $M_{EW}\lesssim 10^2-10^4$ GeV. 
From this point of view, the delicate fine-tuning required to get a small EW scale would be a mirage, since parameters yielding higher EW scales would be in the {\it swampland} and would not count as possible consistent theories. 
This would bring a new perspective into the issue of the EW hierarchy.
 \end{minipage}
\end{center}
\newpage
\setcounter{page}{1}
\pagestyle{plain}
\renewcommand{\thefootnote}{\arabic{footnote}}
\setcounter{footnote}{0}

\section{Introduction}

The deep infrared region of the Standard Model (SM), the region below the electron mass $m_e$ threshold,  is quite 
simple. It only includes a few bosonic degrees of freedom (2 from the photon and 2 from the graviton) and a few
fermionic degrees of freedom (6 or 12 depending on whether they are Majorana or Dirac). The mass scale of neutrinos 
is in the range $m_\nu\simeq 10^{-1}-10^{-2}$ eV (although it is not yet excluded one neutrino being massless). 
The other mass scale relevant in this SM infrared world is the cosmological constant (c.c.) which is 
$\Lambda_0\simeq 3.25\times 10^{-11}\ eV^4 =(0.24\times 10^{-2} eV)^4$.   So it is an experimental fact that 
with a good approximation 
\beq
\Lambda_0 \ \simeq \ (m_\nu)^4  \ .
\label{coinc}
\eeq
It has always been intriguing  the
proximity of the c.c. scale to that of neutrino masses since both scales seem to have a very different origin.
The c.c. comes from the vacuum energy of the SM and its smallness is a major puzzle in the theory. One possible explanation 
of the smallness of the c.c. is the existence of a landscape of vacua, as in string theory \cite{BT,BP}, with this small value required 
for the development of galaxy formation, as first suggested by Weinberg \cite{Weinberg:1988cp}. On the other hand the smallness of neutrino masses
arises naturally from a see-saw mechanism, if neutrinos are Majorana, whereas it is less natural to attain so small 
neutrino masses in the Dirac case.   
It would be interesting to find links between the values of  the $ \Lambda_0^{1/4}$ and $m_\nu$  scales which are quite close 
and around 7 orders of magnitude smaller than the next higher mass scale given by the electron mass $m_e$.

In ref.\cite{ArkaniHamed:2007gg}  Arkani-Hamed et al  explored this deep infrared SM region
by making  the interesting exercise of exploring the possible vacua that
could be obtained by compactifying the SM action to lower dimensions. They found that there is a richness of vacua, a real 
landscape of vacua, both with AdS and dS geometries in 3D and 2D (see  also \cite{Fornal:2011tw,Arnold:2010qz} for
low  $<3D$ SM compactifications). The potential for the moduli of the compactification is induced by the
Casimir effect of the lightest particles of the SM. The existence or not of these lower dimensional vacua turns out to depend 
sensitively on the value of neutrino masses. For example, they found that if all neutrinos are Majorana and we compactify
down to 3 or 2 dimensions  AdS  SM vacua do appear for any values of neutrino masses consistent with experiment.  Interestingly, these vacua 
are almost identical to the 4D SM  at distances larger than 20 microns or so.
Still, these 3D,2D vacua look like a curiosity with no real physical relevance to our world.

In an apparently very unrelated development, there has been in the last few years a renewed interest in the Weak Gravity Conjecture (WGC)
\cite{WGC}.
In  simplified terms the WGC states that in any consistent theory of quantum gravity, the gravitational interaction must be always weaker than 
any other interaction. This statement is motivated by blackhole arguments and has been shown to be a powerful criterium to
determine whether an effective field theory has a consistent UV completion (see \cite{WGC1,WGC2,WGC3} for some recent applications of the
WGC). The main support for the WGC comes from 
 the fact that no contradiction has been found with any string theory example. Recently a {\it sharpened } version of the WGC has 
 been put forward by Ooguri and Vafa \cite{OV} with a quite restrictive corollary (see also \cite{MoreOV} for related work).
 It states that no stable non-SUSY AdS vacua can be consistent with quantum gravity. 
 They also note in passing  that  if the AdS  SM lower dimensional vacua of \cite{ArkaniHamed:2007gg} exist and are stable, then the 4D SM 
 itself could not be completed in the UV. In particular a minimal setting with Majorana neutrinos would be ruled out. 
 
 This is a very interesting remark. The weakest point in the argumentation is that, even if we take for granted this sharpened WGC, it
 applies only if the said 3D,2D vacua are actually stable. If they are unstable, no inconsistency with quantum gravity would appear,
 which would result in no constraints. Some potential instabilities (like  decay into Witten's bubble of nothing) are not present in these SM vacua,
due to the periodic boundary conditions of the fermions. However one may argue that other potential instabilities may appear e.g. 
in the context of a 4D landscape of vacua in string theory, in which tunneling in 4D would have parallel transitions
in lower dimensions, rendering the lower dimensional vacua unstable, and hence leading to no
constraint on neutrino masses or any other physical parameter.

In this paper we reanalyze the issue of the possible constrains on neutrino masses 
but also on the value of the c.c. from the assumption that no lower dimensional AdS vacua of the SM should exist.
We also do this analysis if additional light BSM particles (axions, Weyl fermions, Dirac fermions)  are present well bellow the
electron mass threshold. We are aware that this assumption may be unjustified, since the stability of the 
AdS lower dimensional SM is far from clear, as we will discuss in the text.
Still our knowledge of the 4D landscape of vacua is very poor and their stability is not excluded.
Furthermore, the fact that this assumption leads to intriguing connections between the c.c., the neutrino masses and 
possible additional very light degrees of freedom make this study worthwhile.  In fact we find quite amazing that
a simple very abstract condition like the absence of lower dimensional SM AdS vacua leads to interesting and
potentially testable conditions on the infrared degrees of freedom of the SM. One would have expected that
such abstract condition would had lead to totally wild predictions, and we rather find conditions which are close to be fulfilled
by the SM or some simple extension. In our analysis we confirm that a simplest scheme with 3 Majorana neutrinos would be ruled out within this scheme. 
However the addition e.g. of a single very light Weyl fermion to the SM makes the 
Majorana possibility viable. Dirac neutrinos are viable for the lightest neutrino light enough.
 So e.g. a potential measurement of  (natural hierarchy) Majorana masses at $\nu$-less double $\beta$-decay experiments would 
then imply some additional BSM physics like the existence of additional  very light sterile neutrinos. 

We
also analyze in detail the role of the 4D c.c. on the constraints and find that the 4D c.c. has a  lower bound depending on the
value of neutrino masses.  As the c.c. grows above the neutrino mass scale, the easier is to avoid that AdS vacua develop.
This is important because it is the first argument for a non-vanishing $\Lambda_4$  based only on particle physics and 
not on cosmology.

The bound of the lightest neutrino mass in terms of $\Lambda_4$ allows to a draw another   important conclusion. Indeed, for such  fixed
data this bound implies {\it an upper bound on the EW scale}. This is explicitly seen in fig.~\ref{fig:vevcc} which show how
for  EW scales larger than $10^2-10^4$  GeV   AdS  3D vacua develops,  at fixed cosmological constant.  
From the present perspective the Higgs scale is small compared to the UV scale because of the smallness of the c.c. With values of $\Lambda_4$ 
as observed in cosmology, and reasonable non-vanishing lightest neutrino Yukawa, scales larger than the measured EW scale would yield theories with 3D,2D AdS vacua.  From the  Wilsonian effective field
theory point of view the smallness of the Higgs scale looks like a tremendous fine-tuning. However such a fine-tuning would be a mirage since 
parameters yielding higher Higgs mass scales or vevs cannot be embedded into a consistent theory of quantum gravity and hence do not count 
as possible consistent theories.

The structure of this paper is as follows. In the next section we  briefly review the Weak Gravity Conjecture in connection with AdS vacua 
discussed in \cite{OV}. We also critically discuss the issue of the instability of the AdS vacua obtained upon compactification of the SM
to lower dimensions. In section 3  we study the  3D AdS vacua obtained from the interplay of Casimir forces and the cosmological constant term.
We discuss the limits on neutrino masses obtained imposing the absence of AdS vacua and show how the 4D cosmological constant is 
bounded below by a simple function of neutrino masses, deriving an approximate formula  eq.(\ref{cotilla}). The same analysis but for
toroidal compactifications to 2D is presented in section 4. In section 5 we study how the presence of additional light states beyond the SM ones
modify the previous limits on neutrino masses. The analysis includes the addition of Weyl, Dirac/gravitino and axion states.  In section 6 
we discuss the upper bounds on the EW scale obtained from the absence of AdS 3D vacua.
We present some conclusions and a summary of the results in section 7.

\section{The Weak Gravity Conjecture and AdS vacua}

\subsection{The Ooguri-Vafa conjecture}

The Weak Gravity Conjecture states that, in theories of quantum gravity with a p-form gauge field, there must exist an electrically charged object with charge $Q$ and tension $T$ satisfying
\beq
T\leq M_p^2 Q
\eeq
in order to allow for (sub)extremal black holes to decay and avoid the usual trouble with remnants. In the last years there has been a lot of progress generalising the conjecture for multiple gauge fields and applying it to inflation \cite{WGC1,WGC2,WGC3}. However, a proof has not been found yet, and the strongest evidence for the conjecture in fact comes from the lack of a counter-example in string theory up to now.
 Ooguri and Vafa proposed in \cite{OV} a sharpened version of this conjecture, claiming that the equality can only be satisfied if the charged object is BPS and the theory is supersymmetric. This has dramatic consequences for the AdS/CFT duality as we review in the following. It implies that any non-supersymmetric AdS vacuum supported by fluxes must contain a membrane charged under the flux whose tension is smaller than its charge. 
If this is the case, the possibility of nucleating such a membrane corresponds to an instability of the vacuum, as shown by Maldacena  et al. in \cite{Maldacena:1998uz}
(see also \cite{Barbon:2010gn,Harlow:2010az}).
Once it is nucleated, the bubble will expand and reach the boundary in a finite time, since the electric repulsion wins over the tension of the expanding  bubble, describing the tunneling to a vacuum with a lower value of the flux. Hence, all non-supersymmetric AdS vacua supported by fluxes are at best metastable. In other words, stable non-supersymmetric AdS vacua belong to the swampland, i.e. the set of quantum field theories which are not consistent with quantum gravity and cannot be embedded in the string landscape.

As reviewed in the Introduction, it is known \cite{ArkaniHamed:2007gg}  that three dimensional AdS vacua can appear upon compactifying the Standard Model on a circle. The appearance of these vacua  depends on the value of the neutrino masses with respect to the cosmological constant and its Dirac/Majorana nature. In particular, in the absence of new low energy physics, Majorana neutrinos necessarily give rise to AdS vacua in three dimensions. If these vacua are stable, they would be inconsistent with the above conjecture. As Ooguri and Vafa commented 
in \cite{OV}, this would rule out the possibility of Majorana neutrinos in the SM. Before turning to a more thoughtful analysis of these constraints, let us comment, though, on the issue of stability.

\subsection{Instabilities in the landscape}

The above considerations rely on the assumption that the would-be AdS vacua in three dimensions are stable. However, if the parent four-dimensional deSitter vacuum is unstable, this instability could be inherited by the three-dimensional vacuum wiping out any inconsistency with the above quantum gravity conjectures. This might occur if our four-dimensional vacuum belongs to a landscape of consistent vacua connected by quantum transitions, as suggested by string theory. Then, it would be unstable to tunneling into other parts of the landscape. Unfortunately, our knowledge of the string landscape is very limited and a estimation of the decay rate is completely out of reach at present. We can, though, discuss under what circumstances the four-dimensional instability would also yield an instability in lower dimensions.

Let us assume that the Standard Model lives within a landscape of vacua and that tunneling between different vacua can be described by using semiclassical gravity. Vacuum decay is then described by nucleating a bubble of true vacuum in a region of false vacuum which starts growing approaching the speed of light from the point of view of an outside observer. In deSitter, the bubble radius $R_b$ cannot be larger than the deSitter length $l_4\sim H^{-1}$ (larger bubbles contract instead of expanding). Upon compactification on a circle, the 4d bubble can give rise to a 3d bubble obtained by wrapping the corresponding domain wall on the $S^1$. If the 3d vacuum is deSitter or Minkowski, this bubble will always describe an instability in 3d. However, this is not necessarily the case if the vacuum is AdS. Gravitational effects imply that the radius of a static domain wall is given by the AdS length $R_b\sim l_3$, which means that only smaller bubbles will expand and mediate vacuum decay. In other words, even if the four-dimensional vacuum is unstable, the three-dimensional vacuum will remain stable if the bubble radius in four dimensions is smaller than the $dS_4$ length but still bigger than the $AdS_3$ length, i.e. $l_3<R_b<l_4$.

Let us compute how big is the stability window for the case at hand. The $dS_4$ length scale  in our universe is given by
\beq
l_4=\frac{M_p}{\sqrt{V_0}} \sim  4.8 \cdot 10^{41}\text{ GeV}
\eeq
where we have used that the cosmological constant is $V_0=2.6 \cdot10^{-47} \text{ GeV}^4$. Upon compactifying on a circle of radius $R$, the $AdS_3$ length of the resulting three dimensional space reads
\beq
 l_3=\frac{M_p^{3d}}{\sqrt{V_{R_0}}} \label{l3}
\eeq
where $M_p^{3d}=\sqrt{2\pi R_0}M_p$ and $V_{R_0}$ is the potential energy evaluated at the minimum radius $R_0$. Borrowing the results from next section for $R_0$ and $V_{R_0} $ we can compute the value of $l_3$, obtaining 
\begin{align}
&\text{Majorana NH}\ :\ 4.7 \cdot 10^{39}\leq l_3\leq 1.7 \cdot 10^{41}\text{ GeV} \rightarrow 2.9\lesssim l_4/l_3\lesssim 100\\
&\text{Majorana IH}\ :\ 4.7 \cdot 10^{39}\text{ GeV}\leq l_3\leq 2.6 \cdot 10^{40} \rightarrow 18.2\lesssim l_4/l_3\lesssim 100\\
&\text{Dirac NH/IH}\ :\ l_3\geq 1.2 \cdot 10^{39}\text{ GeV} \rightarrow l_4/l_3\lesssim  410
\end{align}
Notice that the result depends on whether the neutrino particles are Majorana/Dirac with Normal/Inverse hierarchy. The lower bound for $l_3$ comes from the upper limit on the neutrino masses given by Planck 2015, while the upper bound is determined by the lowest neutrino mass which yields an AdS vacuum. As already pointed out in \cite{ArkaniHamed:2007gg}, the stability window is very small if the lightest neutrino is approximately massless. However, it can be made quite large for the case of heavier neutrinos, still consistent with the Planck cosmological bound. In overall, the $AdS_3$ length can vary between zero and two orders of magnitude.
Therefore, instabilities in four dimensions described by a growing bubble whose size is of order  $0.01\, l_4\lesssim R_b\lesssim l_4$ will not yield instabilities in three dimensions. The question now is, in which cases will the membranes mediating vacuum decay have such a critical radius?

Let us first assume that the instability in four dimensions can be described by a Coleman-De Luccia (CDL) instanton within the thin-wall approximation
\cite{coleman}. The size of the bubble is given by
\beq
R_b^{2}=\frac{1}{\left(\frac{\gamma}{\kappa T}\right)^2+\Lambda_i}
\eeq
where $\gamma=\frac{(\kappa T)^2}{4}-\Delta\Lambda_i$ and $\Delta\Lambda=\Lambda_i-\Lambda_f$. We also use the standard notation for the cosmological constant $\Lambda_i=\kappa V_i/3$ with $\kappa=1/M_p^2$. There are two interesting limits depending on whether gravitational effects are important ($T\gg \Delta\Lambda$) yielding $R_b\simeq 4/(\kappa T)$ or negligible ($T\ll \Delta\Lambda$) recovering the flat limit $R_b\simeq \kappa T/\Delta\Lambda$. The case of interest for us, $R_b^2\simeq \Lambda_i^{-1}$, corresponds to an intermediate situation and will happen whenever $\gamma\simeq 0$ leading to
\beq
T^2\simeq 4M_p^4\Delta\Lambda
\label{TV}
\eeq
Since $\Lambda_i$ in our universe is very small, this relation has to be satisfied with a high accuracy. More concretely, if $\epsilon\equiv T^2- 4M_p^4\Delta\Lambda$ one needs $\kappa \epsilon/(4T)\ll\Lambda_i$. As explained above, this is the largest radius the bubble can have in deSitter, and gives rise to a very suppressed tunneling rate at the edge of stability. Intuitively, it corresponds to the case in which the SM is separated from other vacua in the landscape by huge potential barriers. Furthermore, in a supersymmetric theory it corresponds to the BPS bound (static domain walls are given indeed by BPS membranes). 
Since we are in a non-supersymmetric configuration, the membrane action might receive corrections that bring it away from the above bound. If those corrections go in the direction of decreasing the tension $T<M_p^2 Q$ (in a way consistent with the WGC above) and supersymmetry is only slightly broken, one might expect that the condition \eqref{TV} is still approximately satisfied. In such a case, these membranes would fit in the window $l_3<R_b<l_4$ and the 3d vacuum would be stable. However, any quantitative analysis is model dependent and out of reach at present.

On the other hand, such a radius is characteristic of a Hawking-Moss (HM) process
(see  e.g.\cite{KKLT,deAlwis:2013gka,Clifton:2007en,Brown:2010mg,Brown:2011ry} and references therein).
A HM instanton describes the quantum transition of a field starting at the minimum and emerging at the top of the barrier due to quantum fluctuations in a sort of Brownian motion
\cite{Clifton:2007en,Brown:2010mg,Brown:2011ry}. 
After emerging, the field will roll down the potential until the next minimum. This process allows to connect minima for which a CDL instanton does not exist, and has been argued to be essential to populate the landscape \cite{Brown:2011ry},
since up-tunneling from AdS cannot proceed through usual CDL instantons. The decay rate of this stochastic diffusion process is equivalent to that of an homogeneous tunneling of the whole universe in which the bubble radius is $R_b\sim l_4$. A HM process will be dominant with respect to CDL whenever the thickness  of the barrier is bigger than the height. Therefore, if the SM is separated from other vacua in the landscape by thick barriers, the corresponding 3d vacua might be stable. 

Yet another possibility would be that the 4d vacuum is stable, but an instability appears upon compactification. The only known example of this type on a circle compactification is the Witten's bubble \cite{Witten}, which describes the decay of spacetime to nothing. However, this bubble is only topologically consistent with antiperiodic boundary conditions for the fermions around $S^1$, while the $AdS_3$ vacuum exists only for periodic boundary conditions. Therefore, the bubble of nothing is not allowed in our case.

To summarize, it seems that the $AdS_3$ vacua will be stable unless the parent $dS_4$ is unstable and the corresponding bubble size is much smaller than the $dS_4$ length, so it does not lie in the range $l_3<R_b<l_4$. Unfortunately, without a better understanding of the string landscape, we cannot argue one way or another. From now on, we will explore the consequences of assuming that the derived minima are stable. According to the Ooguri-Vafa conjecture, a stable non-supersymmetric AdS vacuum is not consistent with quantum gravity, which leads to interesting constraints on the SM matter spectrum to avoid the appearance of AdS minima upon compactification. We find interesting that the constraints derived in this way are close to the experimental bounds on neutrino masses for the observed value of the c.c.

\section{ AdS Casimir SM vacua in 3D }

The conjecture of Ooguri and Vafa implies that no stable non-SUSY AdS SM vacua should exist. Assuming background independence, this statement should apply to any lower dimensional
compactification of the SM. The simplest case is the 3D in which the SM is compactified on a circle, which we will discuss in this section. 
The compactifications down to 2D are richer, in the sense that there are more options,  the simplest one being the compactification on a 2-torus, which we will study in the next section. Furthermore one can switch on
electromagnetic fluxes through the torus, giving rise to a rich spectrum of vacua.  More generally one can consider compactifications on general Riemann surfaces.
Those have been argued in \cite{ArkaniHamed:2007gg,Arnold:2010qz} not to lead to new vacua. The same has also been argued to be the case of 1D vacua
\cite{Fornal:2011tw}) (quantum mechanics).  For these reasons
we will concentrate in this paper on 3D vacua and 2D toroidal vacua with no fluxes, which are the only vacua in which the Casimir contribution plays an
important role and can lead to constraints on the spectrum of neutrino masses and other BSM very light additional particles.

\subsection{The radion potential in 3D}

In this section we review the origin and numerics of the  3D SM vacua first discussed in \cite{ArkaniHamed:2007gg}. 
The  3D action obtained upon compactification of the SM on a circle of radius $R$ has the form
\begin{eqnarray}
S_{SM+GR}= \int d^3x \sqrt{-g_3} (2\pi r)\left[\frac{1}{2}M_p^2\mathcal{R}_{(3)}-\frac{1}{4}\frac{R^4}{r^4}W_{\mu\nu}W^{\mu\nu}-M_p^2\left(\frac{\partial R}{R}\right)^2 - \frac{r^2\Lambda_4}{R^2}\right].
\label{eq:action}
\end{eqnarray}
Here $M_p$ is the 4D reduced Planck mass, $M_p=(8\pi G_N)^{-1/2}$ and $\Lambda_4$ is the 4D cosmological constant.
The scale $r$ is later to be fixed equal to the vev of the radion $R$.  It also displays the action of
the graviphoton with field strength $W_{\mu\nu}$ and the radion field $R$. The action shows a runaway potential for the radion coming from the
4D cosmological constant
\begin{equation}
V_{\Lambda}(R)=\frac{2\pi r^3 \Lambda_4}{R^2}.
\end{equation}
However the 4D c.c.  is so tiny that the  quantum contribution of the lightest SM modes to the vacuum energy may become important for the 
computation of the radion potential.
The 1-loop corrections to the effective potential of SM particles can be derived from the Casimir energy coming from loops wrapping the circle. 
For massive particles such contributions  are exponentially supressed like   $\propto e^{-2\pi m R}$ for  $R\gg 1/m$. This means that only particles with
mass lighter than $1/R$ contribute significantly.  In the case of massless particles the Casimir contribution to the potential becomes very simple.
One obtains
\begin{equation}
V_C\ =\  \mp \frac{n_0}{720 \pi R^3},
\end{equation}
that is written in the Weyl-rescaled metric as,
\begin{equation}
V_C(R) \ =\ \mp \frac{n_0}{720 \pi}\frac{r^3}{R^6}.
\end{equation}
 The minus sign stands for bosons and the plus sign for fermions with periodic boundary conditions (minus for antiperiodic).
The integer $n_0$ is the number of degrees of freedom of the particle (two for massless vector bosons, two for Majorana fermions, four for Dirac fermions, etc).

\begin{figure}[t]
	\begin{center}
        \includegraphics[scale=0.33]{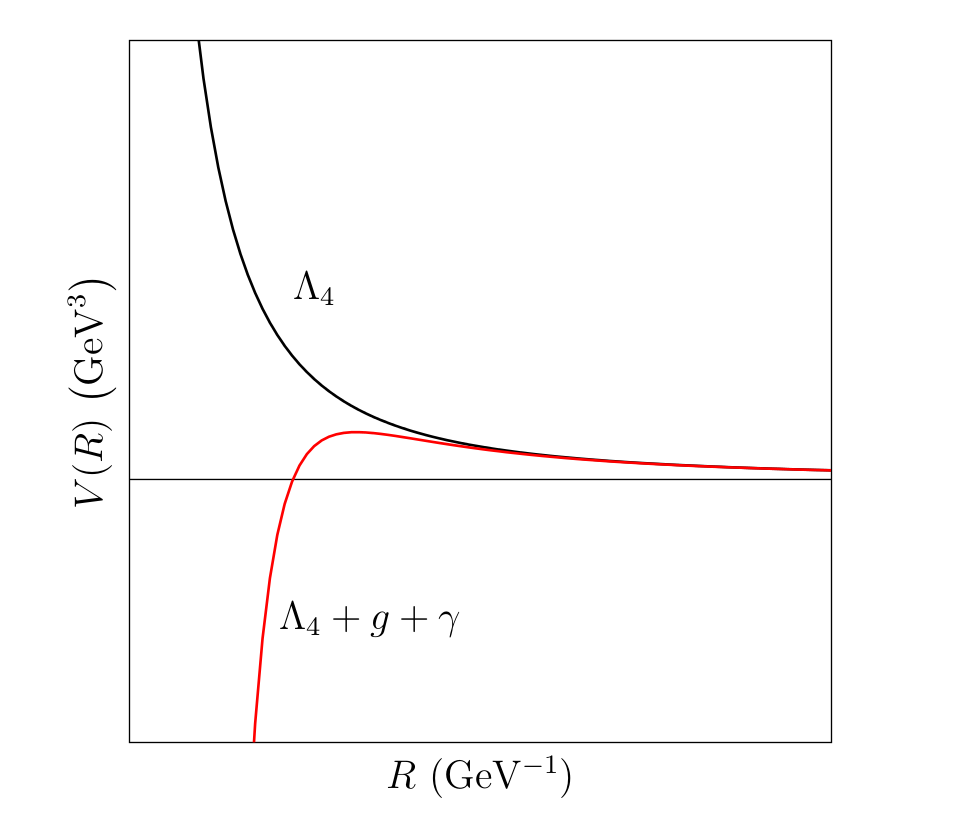}
		\caption{\footnotesize Effective potential as a function of the radion field, R, for the cosmological constant (black) and
		the sum of the cosmological constant, graviton and photon contributions (red).}
		\label{fig:graviphoton}
	\end{center}
\end{figure}

The only massless degrees of freedom in the SM+gravitation are  $4=2+2$ from the photon and the graviton. 
If we only take into account these contributions the effective potential reads,
\begin{eqnarray}
V (R)= \frac{2\pi r^3 \Lambda_4}{R^2} - 4\left(\frac{r^3}{720\pi R^6}\right),
\label{eq:masslesscontri}
\end{eqnarray}
where the number four comes from the sum of the degrees of freedom of the massless particles. In Fig.~\ref{fig:graviphoton} the contributions from the massless states and the cosmological constant are depicted. The contribution of the cosmological constant is shown as a black line.  If we include the massless states, the graviton and the photon, we see that the effective potential, red line, drops for small $R$. In this case there is no minimum. It is clear that the negative sign of the bosonic massless states pushes the effective potential to negative values for small radius due to the sixth power of the radion field, $R^{-6}$ compared with the squared of the cosmological constant, $R^2$, that is important for larger values of $R$. However a maximum appears due to the different behaviour of the two contributions. This
maximum occurs at $R_{max}$ \cite{ArkaniHamed:2007gg},
\begin{equation}
R_{max}=\left(\frac{1}{120\pi^2\Lambda_4}\right)^{1/4}.
\end{equation}
Using the value of the cosmological constant, $\Lambda_4=2.6 \cdot 10^{-47}$ GeV$^4$ \cite{Olive:2016xmw},
\begin{eqnarray}
R_{max}=\left(\frac{1}{120\pi^2\Lambda_4}\right)^{1/4}=7.55\cdot 10^{10}\,{\rm GeV}^{-1},
\end{eqnarray}
and the associated mass scale  here will be,
\begin{eqnarray}
M_{m}=\frac{1}{2\pi R_{max}}=2.11\cdot 10^{-3}\,{\rm eV}.
\end{eqnarray}
Interestingly, this scale is close to the neutrino mass scale. 
As we decrease the value of $R$, the next threshold in the SM is that of neutrino masses. With periodic boundary conditions for neutrinos, 
schematically 
the potential is modified as
\beq
V (R)\ \simeq \  \frac{2\pi r^3 \Lambda_4}{R^2} - 4\left(\frac{r^3}{720\pi R^6}\right) \ +\ 
\sum_{i=\nu_e,\nu_\mu,\nu_\tau} \frac {n_i}{720\pi}\frac {r^3}{R^6} \Theta(R_i-R)  
\label{eq:masslesscontri}
\eeq
with  $R_i=1/m_{\nu_i}$  and $\Theta$ a step function. As $R$ decreases the different neutrino thresholds open and eventually overwhelm  the 
bosonic contribution, giving rise to possible minima,   as long as $R_i < R_{max}$.  Minima turns out to develop   at $R_0\simeq 1/m_\nu$ where
here $m_\nu$ refers to the first threshold for which the number of fermionic degrees of freedom becomes larger than 4.

In practice a correct computation of the minima depends sensitively on the values of the neutrino masses and the cosmological constant, 
and a full computation of the Casimir contributions, including mass effects is required.
In a general case for a particle of mass, $m$ the Casimir energy density is given by  \cite{ArkaniHamed:2007gg}
\begin{eqnarray}
\rho (R)=\mp\sum_{n=1}^{\infty}\frac{2m^4}{(2\pi)^{2}}\frac{K_{2}(2\pi Rmn)}{(2\pi R m n)^{2}},
\label{eq:casimirrho}
\end{eqnarray}
%
where $K_i(x)$ is the Bessel function.  We will use this formula in the computation of the minima below. It is however interesting to expand this formula 
for small  $(mR)$,
\begin{eqnarray}
\rho (R) = \mp \left[\frac{\pi^2}{90 (2\pi R)^4} - \frac{\pi^2}{6(2\pi R)^4}(mR)^2 + \frac{\pi^2}{48(2\pi R)^4}(mR)^4  + \mathcal{O}(mR)^6\right].
\end{eqnarray}
Summing the contributions of the cosmological constant and the particles the effective potential reads,
\begin{eqnarray}
V(R)= \frac{2\pi r^3 \Lambda_4}{R^2} + \sum_i (2\pi R)(-1)^{s_i}n_i\rho_i(R),
\end{eqnarray}
where the sum goes over all the particles in the spectrum, $n_i$ is the number of degrees of freedom of the $i$-th particle
and $s_i=0(1)$ periodic fermions or bosons respectively.
One obtains  a general formula for the potential in terms of the Weyl-rescaled metric for small masses
\begin{eqnarray}
V(R)= \frac{2\pi r^3 \Lambda_4}{R^2} + \sum_i (2\pi R)\frac{r^3}{R^3}(-1)^{s_i}n_i\rho_i(R)\simeq  \notag\\ \simeq
\frac{2\pi r^3 \Lambda_4}{R^2} +  \frac{r^3}{R^6}\frac{\pi^2}{(2\pi)^3}\sum_i (-1)^{s_i}n_i  \left[\frac{1}{90} - \frac{1}{6}(m_iR)^2 + \frac{1}{48}(m_iR)^4\right].
\end{eqnarray}
Setting the scale $r$ such that $2\pi r=1$ GeV$^{-1}$ the effective potential is written,
\begin{eqnarray}
V(R)= 
\frac{( {\rm GeV}^{-3}) \Lambda_4}{(2\pi R)^2} +  \frac{\pi^2 ( {\rm GeV}^{-3})}{(2\pi R)^6}\sum_i (-1)^{s_i}n_i  \left[\frac{1}{90} - \frac{1}{6}(m_iR)^2 + \frac{1}{48}(m_iR)^4\right].
\label{eq:effpotentialapprox}
\end{eqnarray}
Note that this formula is not a good approximation to study the  minima  generated by neutrinos because, as we said, the minima are obtained at $R_0\simeq 1/m_\nu$ 
and hence $(Rm)$ is not in general small.  However in the case of the lightest neutrino (or some additional very light BSM state) $(Rm)$ may be small enough so that 
the dependence on these masses is adequately described by this expression. We will also use it as an inspiration to fit the curve which parametrises the lowest value of the cosmological constant required to get positive vacuum energy in section 3.5.

\subsection{Limits on neutrino masses}
\label{sec:neutrinos}

\begin{figure}[t]
	\begin{center}
        \includegraphics[scale=0.33]{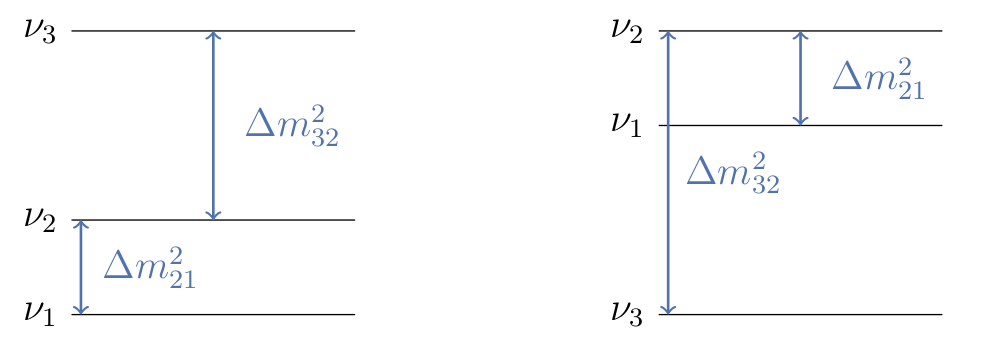}
		\caption{\footnotesize Normal and inverted hierarchies of neutrino masses.}
		\label{hierarchy}
	\end{center}
\end{figure}

As we have discussed in the previous chapter, we  want to impose that no stable AdS vacua of the SM should exist. 
Note that only compactifications  with periodic boundary conditions for the neutrinos are dangerous, since only in this case the 
Casimir energy for fermions is positive.
The existence or not of these
vacua depends also sensitively on the specific values of neutrino masses. 
At the moment  we do not know the absolute masses of the neutrinos, nonetheless we are able 
to measure the difference in masses between them. According to the PDG \cite{Olive:2016xmw} the atmospheric and solar difference in masses are,
\begin{eqnarray}
\Delta m_{21}^2 = (7.53\pm 0.18)\times 10^{-5}\,\, {\rm eV}^2,\\
\Delta m_{32}^2 = (2.44\, \pm 0.06)\times 10^{-3}\,\, {\rm eV}^2 \,\,{\rm (NH)},\\
\Delta m_{32}^2 =(2.51\pm 0.06)\times 10^{-3}\,\, {\rm eV}^2 \,\,{\rm (IH)}.
\end{eqnarray}
We also do not know the nature of the hierarchy of masses, either Normal Hierarchy (NH) , with $m_{\nu_1}<<m_{\nu_2} << m_{\nu_3}$  or
Inverted Hierarchy (IH), with $m_{\nu_1}< m_{\nu_2} >> m_{\nu_3}$ (see fig.(\ref{hierarchy}).  In the NH case, for $m_{\nu_1}\ll m_{\nu_2}$ one gets approximately

\beq
m_2\ \simeq \ 8.6 \times 10^{-3}\ {\rm eV} \ \ ;\ \ m_3 \ \simeq \ 4.9\times \ 10^{-2}\ {\rm eV}
\eeq
The lightest neutrino may be arbitrarily light from these data.  In the case of the inverted hierarchy one has
\beq 
m_{\nu_1}\ \simeq \ m_{\nu_2} \  \simeq  \ 4.9\times 10^{-2} \ {\rm eV}
\eeq
with $m_{\nu_3}$ arbitrarily light.
Using these experimental data, we will constraint the possible values of the lightest neutrino in both NI and IH arising from
the above WGC motivated constraint, the absence of stable AdS vacua \footnote{It has been  recently claimed a slight preference for 
the natural hierarchy from the combined analysis of neutrino data \cite{Simpson:2017qvj,Agostini:2017jim,Schwetz:2017fey}. However the evidence is still very weak \cite{Vagnozzi:2017ovm}.} We discuss the cases of Majorana and Dirac neutrinos in turn.

\begin{figure}[t]
	\begin{center}
        \includegraphics[scale=0.33]{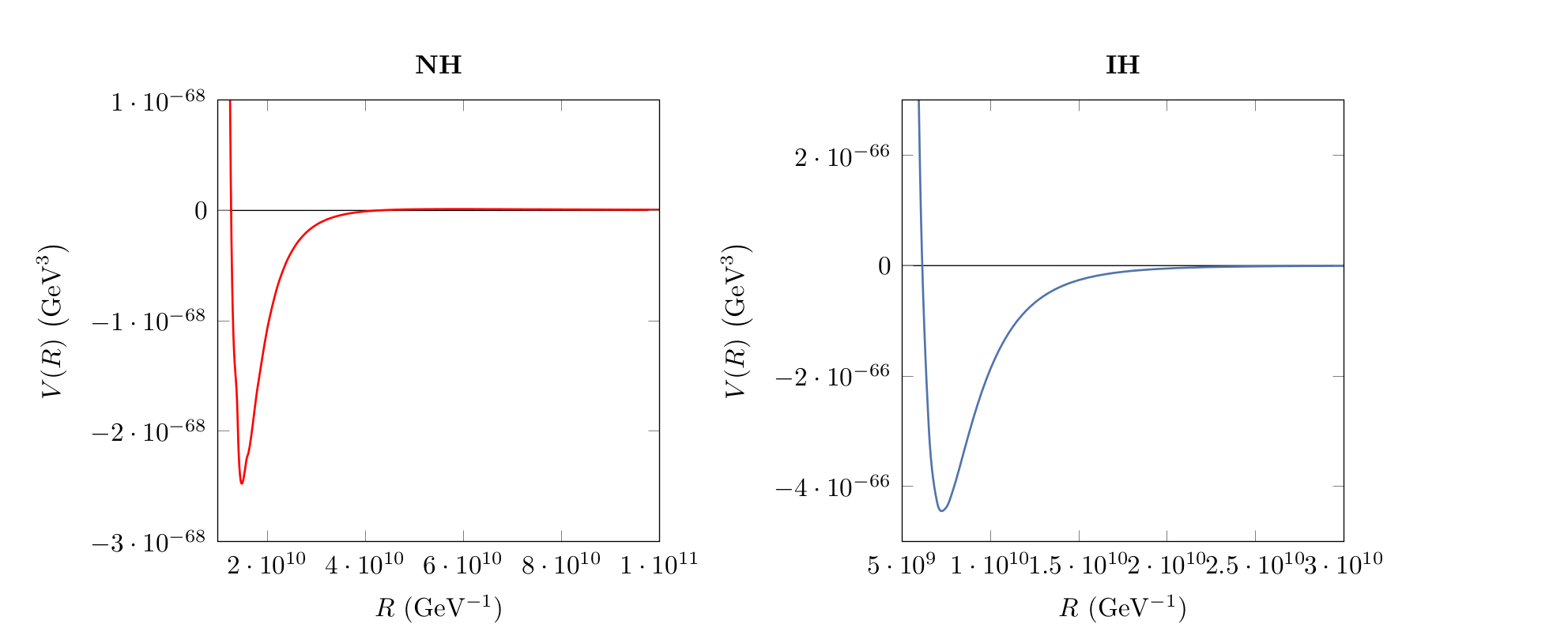}
		\caption{\footnotesize Effective radion potential for Majorana neutrinos when considering normal hierarchy (left) and inverted hierarchy (right). In both cases the lightest neutrino is considered massless, $m_{\nu_1}=0$  for NH and $m_{\nu_3}=0$ for IH.}
		\label{fig:neutrinosmajorana}
	\end{center}
\end{figure}
\begin{figure}[t]
	\begin{center}
        \includegraphics[scale=0.33]{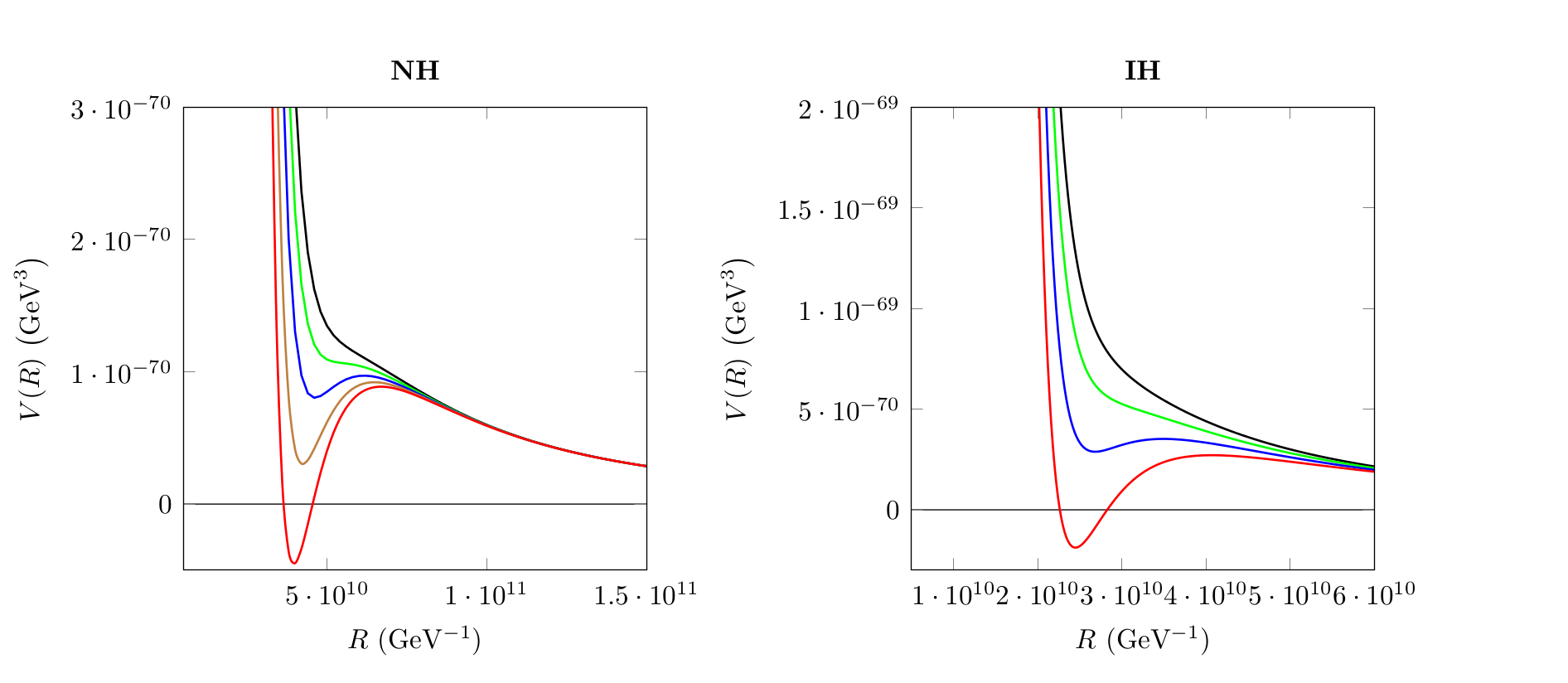}
		\caption{\footnotesize Radion effective potential for Dirac neutrinos when considering normal hierarchy (left) and inverted hierarchy (right). For the case of NH the different lines correspond to several values for the  lightest neutrino mass: $m_{\nu_1}=$ 6.0 meV (black), 6.5 meV (green), 7.0 meV (blue), 7.5 meV (brown) and 8.0 meV (red). In the case of IH the different colours correspond to the lightest neutrino masses: $m_{\nu_3}=$ 1.5 meV (black), 2.0 meV (green), 2.5 meV (blue), 3.0 meV (red).}
				\label{fig:neutrinosdirac}
	\end{center}
\end{figure}

\subsection{Majorana neutrinos}

In the case of Majorana neutrinos we have 6 fermionic and 4 bosonic degrees of freedom, so one expects that AdS vacua will develop. 
In Fig.~\ref{fig:neutrinosmajorana} the effective potential for Majorana neutrinos is shown where the lightest neutrino has a zero mass. On the left panel of Fig.~\ref{fig:neutrinosmajorana} it is assumed a NH for the neutrinos masses where on the right panel of Fig.~\ref{fig:neutrinosmajorana} it is assumed an IH. An AdS vacuum is always formed for this configuration in both hierarchies.  If the mass of the lightest neutrino is different from zero, the mass terms of the potential make the potential deeper.  
So the case of the pure SM with Majorana neutrino masses would be ruled out, as already advanced in  \cite{OV}.

\subsection{Dirac neutrinos}

In Fig.~\ref{fig:neutrinosdirac} the case of Dirac neutrinos is presented. On the left panel of Fig.~\ref{fig:neutrinosdirac} the NH is assumed. In this case a few values for the 
lightest neutrino masses are taken: 6.0 meV (black), 6.5 meV (green), 7.0 meV (blue), 7.5 meV (brown) and 8.0 meV (red). In this case we find different solutions in the effective potential depending on the neutrino masses. For masses greater than 7.73 meV an AdS vacuum is formed, while for masses between 6.7 meV and 7.73 meV a dS vacuum is obtained. In the case where the lightest neutrino is smaller than 6.7 meV there is no vacuum. On the right panel of Fig.~\ref{fig:neutrinosdirac} the IH is assumed. In this case the different colours correspond to the lightest neutrino mass: $m_{\nu_3}=$1.5 meV (black), 2.0 meV (green), 2.5 meV (blue), 3.0 meV (red). For this mass hierarchy we found that for a mass of the lightest neutrino greater than 2.56 meV an AdS vacuum is formed. A dS vacuum is achieved for masses between 2.56 meV and 2.1 meV, and if the lightest neutrino mass is smaller than $m_{\nu_3}<2.1$ meV no vacua is formed. A summary of the masses for which the different vacua are formed is found in Tab.~\ref{tab:dirac}.

\begin{table}
\begin{center}
\begin{tabular}{c | c | c |}
 & NH & IH \\
 \hline
No vacuum & $m_{\nu_1}< 6.7$ meV & $m_{\nu_3}< 2.1$ meV \\
dS$_3$ vacuum & 6.7 meV $<m_{\nu_1}< 7.7$ meV& 2.1 meV $<m_{\nu_3}<$ 2.56 meV \\
AdS$_3$ vacuum & $m_{\nu_1}> 7.7$ meV & $m_{\nu_3}> 2.56$ meV \\
\hline
\end{tabular}
\caption{Ranges of masses of Dirac neutrinos for  different vacua configurations.}
\label{tab:dirac}
\end{center}
\end{table}

\subsection{Cosmological constant versus neutrino masses from 3D vacua}

\begin{figure}[t]
	\begin{center}
        \includegraphics[scale=0.33]{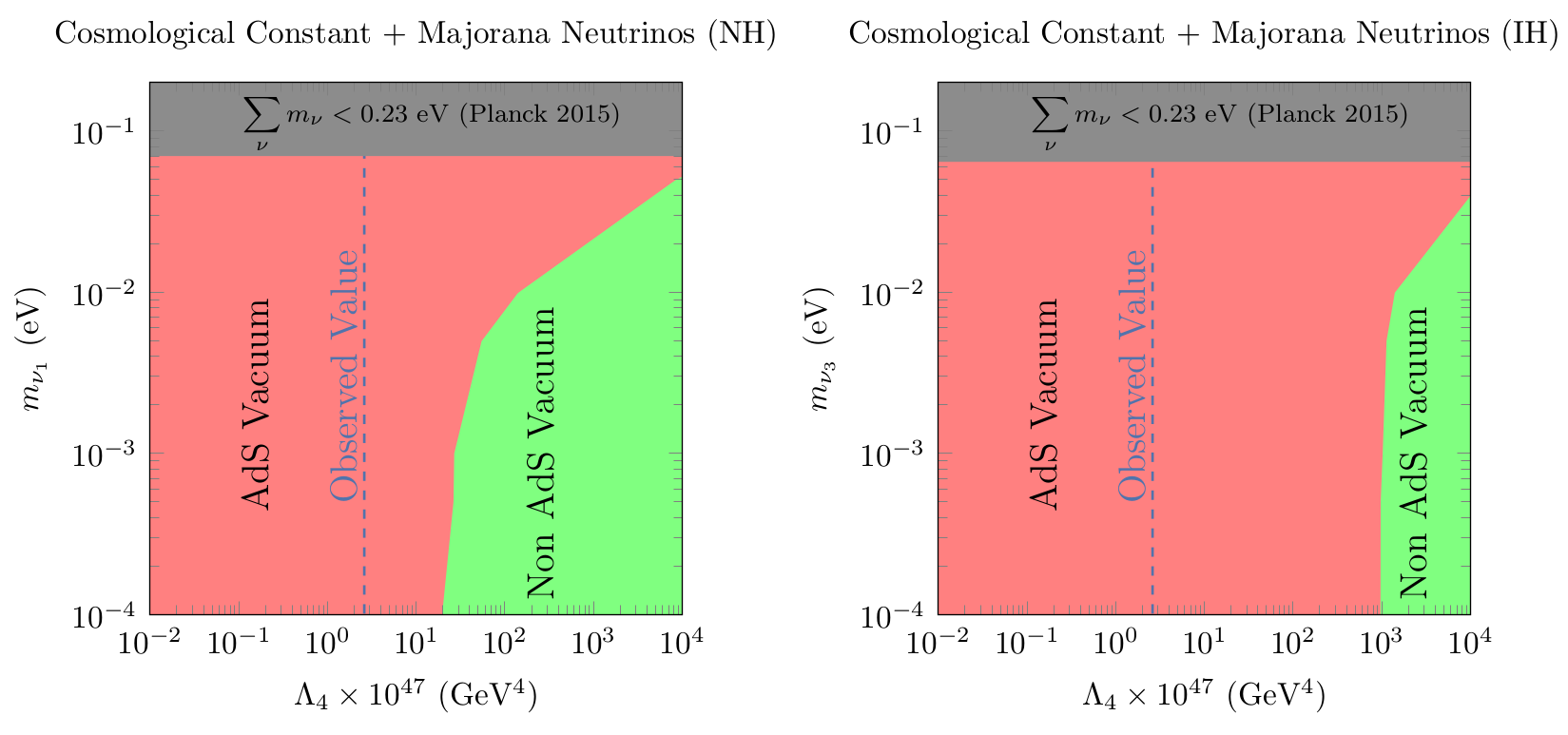}
		\caption{\footnotesize  Majorana neutrinos. Lower bound on the value of the 4D cosmological constant as a function of the
		lightest neutrino mass coming from absence of AdS vacua.  Left: NI. Right: IH.}
		\label{majcc}
	\end{center}
\end{figure}

It is important to remark that the above results depend sensitively on the value of the 4D cosmological constant.
It is clear that, the higher the value of the 4D c.c., the easier becomes to avoid unwanted AdS vacua.  For given neutrino masses,
there is a lower bound on  $\Lambda_4$ coming from absence of AdS vacua.  
To show this dependence we present in fig.\ref{majcc} the allowed values of the lightest neutrino versus the value of the
cosmological constant, both for NH and IH, in the Majorana case.  The areas in red correspond to AdS vacua and should be forbidden. We 
see that in the NI case this bound is around 7 times higher than the experimental $\Lambda_4$ and several orders of magnitude higher in the IH case.
That is why in the Majorana case is impossible to avoid AdS vacua.  We will see later however, that the addition of e.g. just an additional light Weyl fermion to the SM
it is enough to make viable the Majorana neutrino case.

\begin{figure}[t]
	\begin{center}
        \includegraphics[scale=0.33]{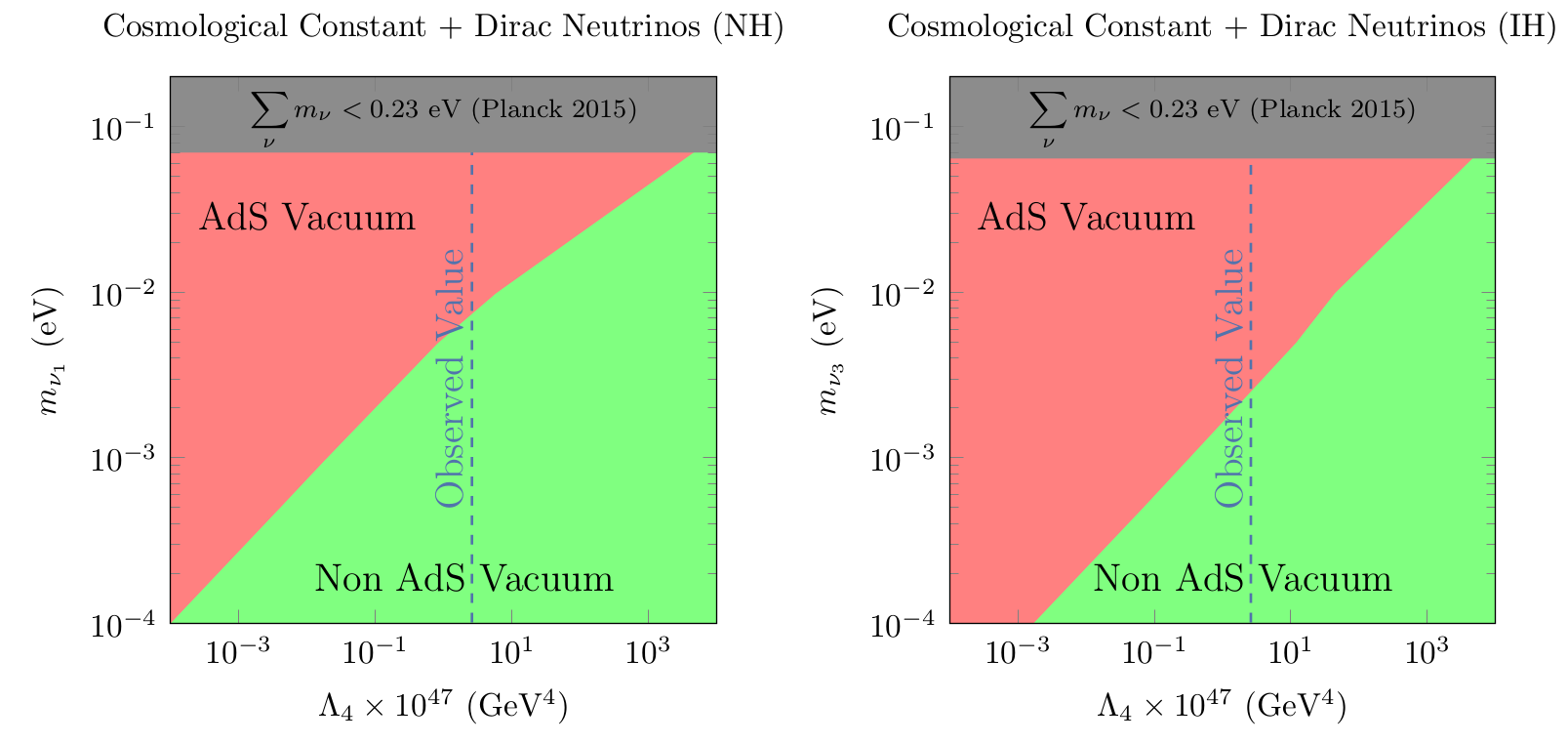}
		\caption{\footnotesize  Dirac neutrinos. Lower bound on the value of the 4D cosmological constant as a function of the
		lightest neutrino mass coming from absence of AdS vacua.  Left: NI. Right: IH.}
		\label{dircc}
	\end{center}
\end{figure}

In the case of Dirac neutrinos the situation is different due to the fact that the number of fermionic degrees of freedom doubles and 
they may pull up the potential before an AdS vacuum can develop. We show in fig.\ref{dircc} the corresponding plot for the
Dirac case.  We see that for any given value of the lightest neutrino, there is a lower bound on the value of the 4D cosmological constant.
For the value of the experimental cosmological constant  one obtains a lower bound on the value of the lightest neutrino mass,
$m_{\nu_1}> 7.7\times 10^{-3}$ eV  for NH and   $m_{\nu_3}> 2.56\times 10^{-3}$ eV for IH.

The form of the curves in figs. \ref{majcc},\ref{dircc}   may be understood from the aproximate equations given in \ref{eq:effpotentialapprox}. Let us assume for the moment that $mR$ is small for the three neutrinos, so the formula \ref{eq:effpotentialapprox} is a good approximation. We can then minimize the potential to get $R_{min}$ and derive the value of $\Lambda_4$ in terms of the neutrino masses for which $V(R_{min})\geq 0$, obtaining
\beq
\Lambda_4\geq\frac{n_f (30 n_f (\Sigma m_i^2)^2 + (4-3 n_f) \Sigma m_i^4) }{(-3072 + 2304 n_f) \pi^2}\label{lambda4}
\eeq
where $n_f=2,4$ for Majorana/Dirac respectively. Above this value for $\Lambda_4$ no AdS minimum is formed. Unfortunately, $mR\sim 1$ for the neutrino masses (specially the heaviest one), so this curve does not fit well enough the numerical results in figs.\ref{majcc} and \ref{dircc}. In fact, the appearance of a minimum is based on a delicate interplay between the contribution from neutrino particles and cosmological constant, so the results are very sensitive to the exact value of these contributions. The inclusion of higher order terms in \ref{eq:effpotentialapprox} does not lead to a notable improvement before the minimization analyses of the potential becomes untractable. However, we can use \eqref{lambda4} as an educated  anstaz to fit the curve which separates the region of AdS and Non-AdS vacuum in the above figures. Assuming that the leading terms will still be given by functions of $(\Sigma m_i^2)^2$ and $\Sigma m_i^4$ but higher order corrections may modify the coefficients in \eqref{lambda4}, we fit our results to the curve
\beq
\Lambda_4\geq\frac{a(n_f) 30 (\Sigma m_i^2)^2-b(n_f,m_i) \Sigma m_i^4 }{384\pi^2}
\label{cotilla}
\eeq
obtaining $a(n_f)=0.184(0.009)$ and $b(n_f,m_i)=5.72(0.29)$ for NH or $b(n_f,m_i)=7.84(0.55)$ for IH, respectively for Majorana(Dirac) neutrinos.
This curve describes quite accurately the numerical results and displays the lower bound on the cosmological constant required to guarantee the absence of AdS vacuum. Interestingly, this bound scales as $m_\nu^4$, as observed experimentally.

The mere existence of these lower bounds is  interesting, since the only input is the value of the 4D c.c., yet the values obtained 
are close to expectations from particle physic models.  Furthermore they give us a rationale as to why a non-vanishing value of the 
c.c. would be expected on arguments not based at all on the need for dark energy in cosmology. The existence of dark energy could have been predicted 
on the basis of these arguments.

\section{AdS Casimir SM vacua in 2D}

%

In the previous section it was shown the 3D compactification of the SM. One can also compactify to 2D \cite{ArkaniHamed:2007gg,Arnold:2010qz}. In this case there are more SM compactifications than in the 3D case. The most simple case is the compactification in a 2D torus, and this is the case that we will follow in the rest of the work. However several compactifications in different manifolds are also possible. One example is the 2D sphere. In this case there is an extra classical contribution from the curvature to the potential. As it was shown in  \cite{Arnold:2010qz} one can only obtain new stable vacuum when magnetic fluxes are switched on, and in this case the 
Casimir contribution of neutrinos becomes irrelevant for those vacua, and no further constraints are obtained.
 For the case of other Riemann surfaces there are no 2D SM vacua configurations even if there are magnetic fluxes.  For those reasons we study in this section the case of the 2D torus compactification with no fluxes, which is the only 2D vacua depending strongly on the neutrino spectrum.

\subsection{The radion potential in 2D. 2D SM vacua and neutrino masses}

As pointed out in Ref.~\cite{ArkaniHamed:2007gg} and then studied  in detail in Ref.~\cite{Arnold:2010qz} the 2D potential can be written as

\begin{eqnarray}
V(a,\tau)=(2\pi a)^2 \Lambda_4 + \sum_{a}(-1)^{F_a}n_aV_{2D-C}^{(1)}[a, \tau_1, \tau_2, m_a],
\end{eqnarray}
with $V_{2D-C}^{(1)}[a, \tau_1, \tau_2, m_a]$ defined by

\begin{align}
 V_{2D-C}^{(1)}[a, \tau_1, \tau_2, m_a]=-\frac{1}{(2\pi a)^2}\left[\frac{2(am)^{3/2}}{\tilde{\tau}^{1/4}}\sum_{p=1}^{\infty}\frac{1}{p^{3/2}}K_{3/2}(2\pi a p m_a \sqrt{\tilde{\tau}_2})\right. & \notag\\
 + 2\tilde{\tau}_2(am)^2 \sum_{p=1}^{\infty}\frac{1}{p^2}K_2\left(\frac{2\pi a p m_a}{\sqrt{\tilde{\tau}_2}}\right) & \notag\\
\left.  +4\sqrt{\tilde{\tau}_2}\sum_{n,p=1}^{\infty}\frac{1}{p^{3/2}}(n^2+ \frac{(am)^2}{\tilde{\tau}_2})^{3/4}\cos(2\pi p n \tilde{\tau}_1)K_{3/2}(2\pi p \tilde{\tau}_2\sqrt{n^2+\frac{(am)^2}{\tilde{\tau}_2}}) \right], &
\end{align}
where the 2D torus is parametrized as 
\begin{eqnarray}
t_{ij}=\frac{a^2}{\tau_2}\left(
\begin{array}{c c}
1 & \tau_1 \\
\tau_1 & |\tau|^2
\end{array}
\right).
\end{eqnarray}
and ${\tilde \tau}_i=\tau_i/|\tau|^2$. In the following we will assume $|\tau|=1$ for the torus.

The minima of the effective potential corresponding to AdS vacua are those verifying the  conditions \cite{ArkaniHamed:2007gg,Arnold:2010qz}
\begin{eqnarray}
V(a, \tau)=0,\quad \partial_{\tau_{1,2}}V(a,\tau)=0,\\
\partial_aV(a,\tau)<0, \partial^2_{\tau_{1,2}}V(a,\tau)>0.
\end{eqnarray}

However not every configuration of the complex structure of the torus leads to the appearance of extrema in the potential. The Casimir contributions to the energy density are invariant under $SL(2,\mathds{Z})$ modular transformations. For that reason only in stationary points of the complex structure one can find extrema of the potential \cite{Arnold:2010qz}. These stationary points in the case of the 2D torus are $\tau = 1$ and $\tau = 1/2 + i\sqrt{3}/2$. As it was pointed out in Ref.~\cite{Arnold:2010qz} only the latter point presents a minimum of the potential. Thus, in the rest of the paper we assume this structure for the 2D torus in the computations. It is important to note that in this case either an AdS vacuum appears or there is no vaccuum at all. 
This scenario is different compared to the 3D case since in the latter also dS vacua could appear.

\subsection{Majorana neutrinos}

\begin{figure}[t]
	\begin{center}
        \includegraphics[scale=0.33]{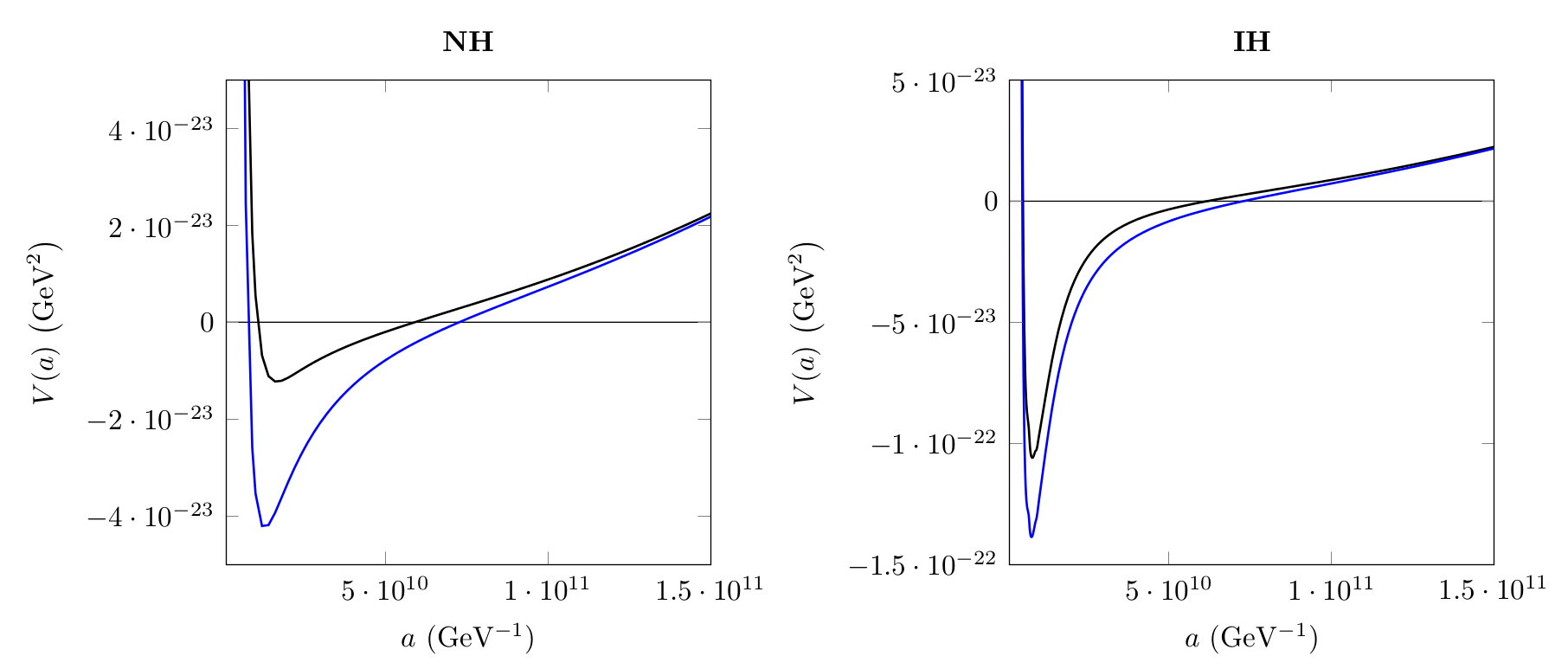}
		\caption{\footnotesize Effective potential for Majorana neutrinos for the 2D case when considering normal hierarchy (left) and inverted hierarchy (right). In both cases the lightest neutrino is considered massless, $m_{\nu_1}=0$ eV for NH and $m_{\nu_3}=0$ for IH (black line) and a mass of $m_{\nu_1}=10^{-2}$ eV for NH and  $m_{\nu_3}=10^{-2}$ eV for IH (blue line).}
		\label{fig:neutrinosmajorana2d}
	\end{center}
\end{figure}

As it was discussed in the case of the 3D compactification,  we expect the presence of AdS vacua due to the fact that there is a bigger number of fermionic degrees of freedom compared to  the bosonic ones. In Fig.~\ref{fig:neutrinosmajorana2d} we show the potential for Majorana neutrinos. Left panel of Fig.~\ref{fig:neutrinosmajorana2d} corresponds to a NH ordering  and the right panel to an IH. As in the case of 3D compactification an AdS vacuum always develops. For a massless lightest neutrino, the black line in both panels of Fig.~\ref{fig:neutrinosmajorana2d}, an AdS vacuum is found which  means that for larger masses this vacuum remains. This can be seen in terms of the blue line depicted in both panels which corresponds to a lightest neutrino mass of $m_{\nu_1}=10^{-2}$ eV for NH and $m_{\nu_3}=10^{-2}$ eV for IH. As in the 3D case the Majorana neutrino contributions 
drive down the potential to an AdS vacuum, and this possibility would be excluded, if no extra light particles are added.

\subsection{Dirac neutrinos}

\begin{figure}[t]
	\begin{center}
        \includegraphics[scale=0.33]{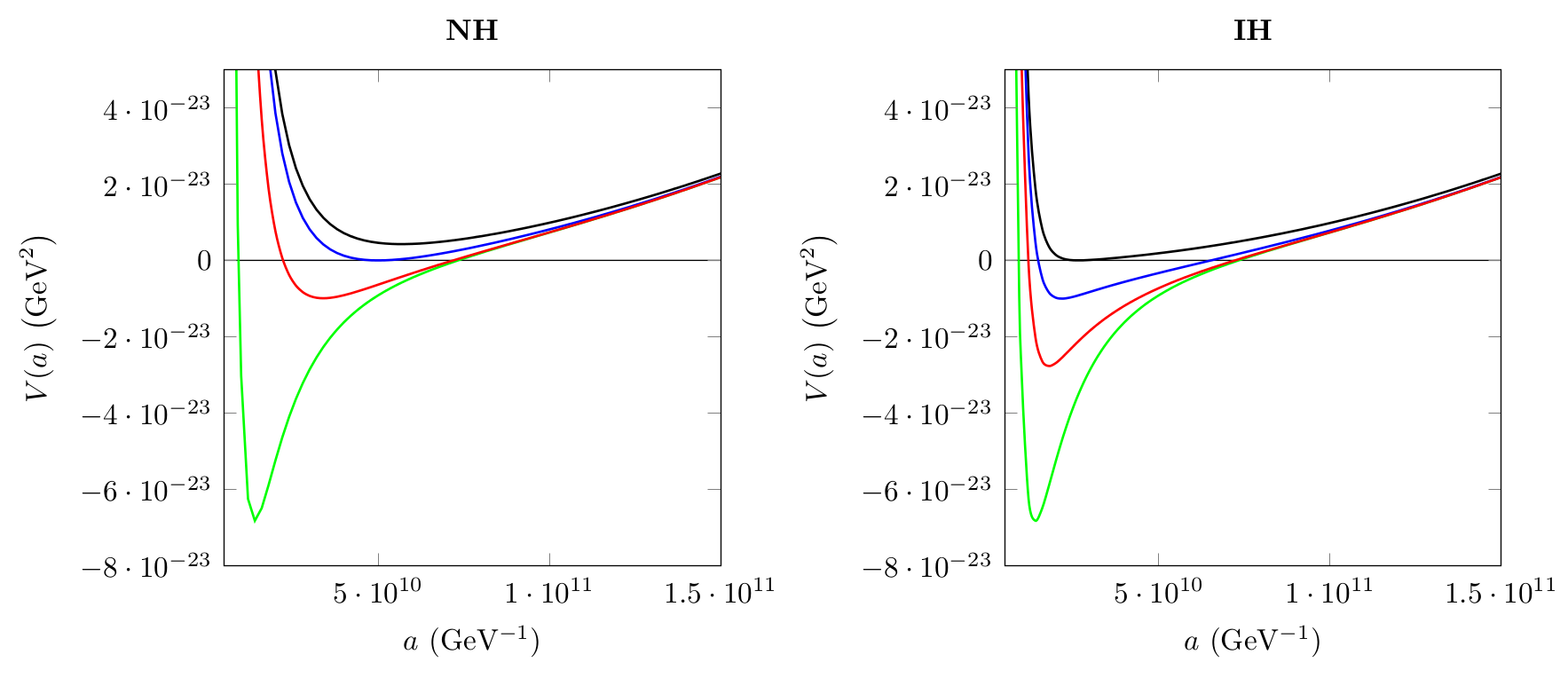}
		\caption{\footnotesize Effective potential for Dirac neutrinos for the 2D case when considering normal hierarchy (left) and inverted hierarchy (right). For the case of NH the different lines correspond to different lightest neutrino mass: $m_{\nu_1}=$ 1.0 meV (black), 4.12 meV (blue), 10.0 meV (red) and 20.0 meV (green). In the case of IH the different colours correspond to the lightest neutrino masses: $m_{\nu_3}=$ 1.0 meV (black), 5.0 meV (blue), 10.0 meV (red), 20.0 meV (green).}
				\label{fig:neutrinosdirac2d}
	\end{center}
\end{figure}

In the case of Dirac neutrinos, one could expect the same  possibilities that we found for the 3D case. In this 2D case however  we can conclude 
immediately when an AdS vacuum is present since there are not dS vacua as it was mentioned before. In Fig.~\ref{fig:neutrinosdirac2d} the potential for Dirac neutrinos is depicted. On the left panel of Fig.~\ref{fig:neutrinosdirac2d} a NH is assumed for the neutrinos while on the right panel an IH is assumed. Different lines represent different neutrino masses: 1.0 meV (black), 5.0 meV (blue), 10.0 meV (red), 20.0 meV (green). In the case of NH (left panel of Fig.~\ref{fig:neutrinosdirac2d}) for masses of the lightest neutrino greater than $m_{\nu_1}\geq 4.12$ meV an AdS vacuum is developed while for masses lighter than that value there is no vacuum at all. In the case of IH (right panel of Fig.~\ref{fig:neutrinosdirac2d}) the mass of the lightest neutrino must be greater than $m_{\nu_3}\geq 1.0$ meV in order for an AdS vacuum to be created. These limits could be compared directly with the ones of Table~\ref{tab:dirac} for the 3D case. In this case the 2D compactification imposes stringent bounds setting lower masses for the mass of the lightest neutrino that induces an AdS vacuum.

\begin{table}
\begin{center}
\begin{tabular}{c | c | c |}
 & NH & IH \\
 \hline
No vacuum & $m_{\nu_1}< 4.12$ meV & $m_{\nu_3}< 1.0$ meV \\
AdS$_3$ vacuum & $m_{\nu_1}> 4.12$ meV & $m_{\nu_3}> 1.0$ meV \\
\hline
\end{tabular}
\caption{Ranges of masses of Dirac neutrinos where different vacua configurations are shown for a 2D torus compactification.}
\label{tab:dirac2d}
\end{center}
\end{table}

\subsection{Cosmological constant versus neutrino masses from 2D vacua}

\begin{figure}[t]
	\begin{center}
        \includegraphics[scale=0.33]{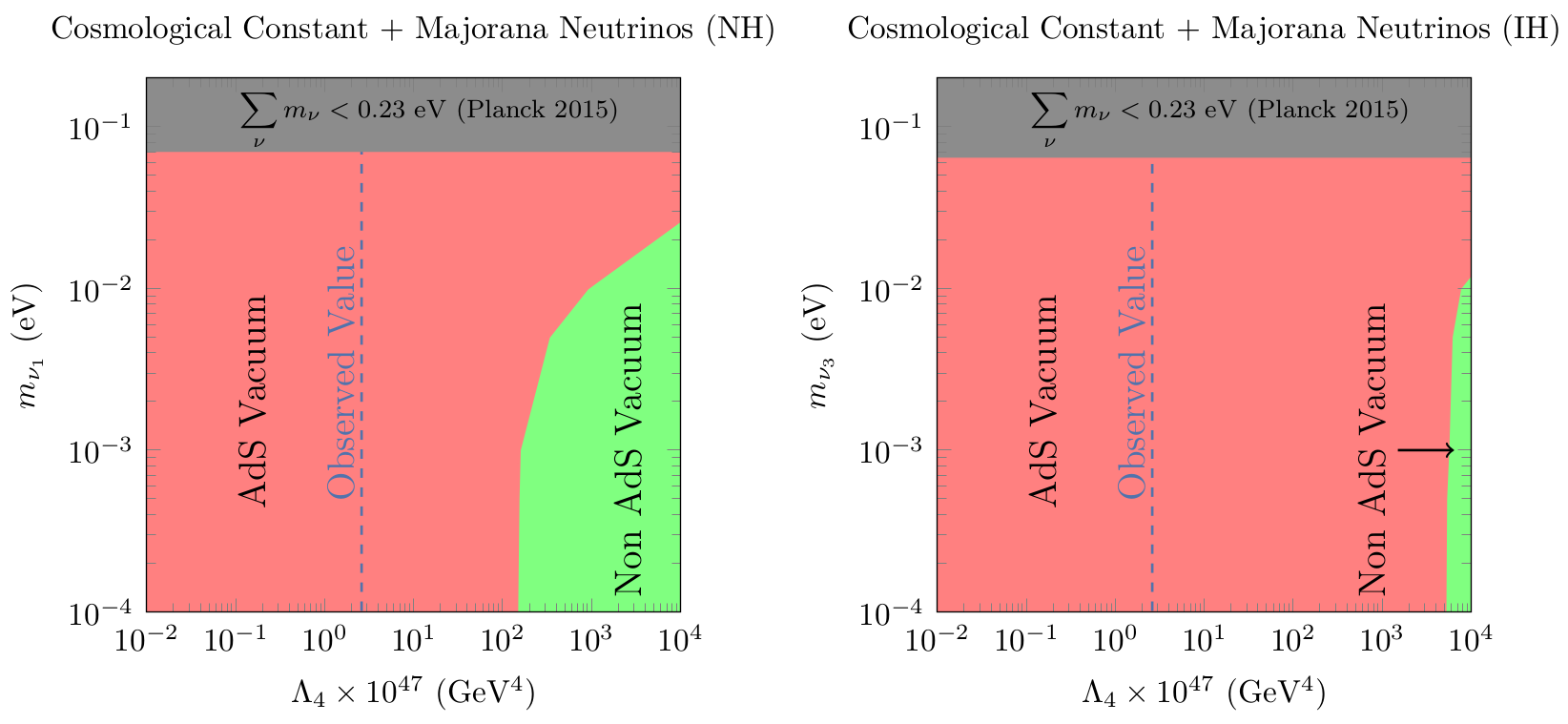}
		\caption{\footnotesize  Majorana neutrinos. Lower bound on the value of the 4D cosmological constant as a function of the
				lightest neutrino mass coming from absence of AdS vacua for 2D compactification.  Left: NI. Right: IH.}
				\label{fig:ccmajorananeutrino2d}
	\end{center}
\end{figure}

As we did in the 3D compactification case,  we can compute how the 4D cosmological constant value could affect the creation of AdS vacua. In Fig.~\ref{fig:ccmajorananeutrino2d} the lower bound on the cosmological constant as a function of the lightest neutrino mass is depicted for the case of Majorana neutrinos. The red area corresponds to AdS vacua and so it is  excluded. The left panel of Fig.~\ref{fig:ccmajorananeutrino2d} shows a NH for  Majorana neutrinos and the right one for  IH. In comparison with the 3D case we see that the limits are more stringent. For the NH scenario we have that the lower value for the cc to have a non-AdS vacuum is 60 times larger than the cc, while in the 3D case this number was 7. This is also de case for IH where now the limits on the minimal value of the cc are one order of magnitude larger.  For the Dirac case  something  similar  happens as we can deduce from Table~\ref{tab:dirac2d}. Note that still  Majorana neutrinos  are  excluded by the observed value of the cc. In order to avoid an AdS vacuum for Majorana neutrinos one would have
 needed  a greater value of the cc than the one observed.

\section{Beyond the SM: adding light fermionic  and bosonic degrees of freedom}

The presence of additional very light particles with masses of the order of neutrino masses or smaller 
can substantially modify the structure or the very existence of 3D or 2D vacua and modify 
or eliminate the bounds above.  Here  we will discuss in turn the addition of extra fermionic or bosonic degrees of freedom separately.
There are of course more complicated possibilities with e.g. additional fermions and bosons at the same time, which can equally be studied
using the above equations.
The effect of the different possibilities for BSM scenarios involving these extra light states is summarized later in table (\ref{tab3D2D}).

\subsection{Adding Weyl, Dirac fermions or gravitino}

The existence of very light neutral fermionic degrees of freedom have been advocated for several  purposes. Some examples are
as follows:

 {\it Sterile neutrinos.}
These particles have been introduced as a generalization of the SM neutrino system (see e.g.\cite{sterile} for reviews). One original motivation was
the presence of such states with a mass of order 1eV 
to account for some neutrino  oscillation anomalies detected at LSND and other neutrino  experiments.
But, more generally,   the presence of sterile neutrinos has been considered for a variety of purposes.
Axinos (SUSY partners of the axion) may also be considered in this class.
Although in specific models sterile neutrinos have masses typically of order 1 eV, very light sterile neutrinos with masses e.g. $m^2 = 6\times  10^{-3}\  eV^2$, relevant for 
Casimir energies,  are also possible (see e.g. \cite{sterilelight}).

{\it Light gravitinos.}
Very light gravitinos appear in models of low scale gauge mediation.  Minimal models of gauge mediation have gravitino masses of
order
\begin{eqnarray}
m_{3/2}=\xi\frac{F}{\sqrt{3}M_p}=\xi \left(\frac{\sqrt{F}}{100\,{\rm TeV}}\right)^2 \times 2.4 \,{\rm eV},
\end{eqnarray}
here $\xi=F_0/F$, where $F_0$ is the fundamental SUSY-breaking auxiliary field scale and $F$ is the spurion auxiliary field in $X=M+\theta^2 F$ . 
This auxiliary field coupled to the MSSM  may be smaller than $F_0$. So, \textit{e.g.} for $\xi=1$ and $F=(10\,{\rm TeV})^2$ one has $m_{3/2}\simeq 2.4 \times 10^{-2}$ eV,
well in the region relevant for Casimir potentials.
There are cosmological upper bounds on stable gravitinos (see e.g.\cite{Oyama:2016lor,Osato:2016ixc,Brust:2013xpv,Pospelov:2010hj} and references therein).
From CMB measurements  one gets $m_{3/2}\leq 4.7$ eV \cite{Osato:2016ixc} and from primordial nucleosintesis  
$m_{3/2}\leq 16$ eV. In gauge mediation there are lower bounds on the gravitino mass coming from consistency with the measured Higgs mass, which gives a lower bound 
on the SUSY breaking scale. Lower bounds depend on the GMSB version.  In  minimal GMSB  one gets $m_{3/2} > 300 eV$ but more general GMSB models 
may yield gravitino masses as low as 1 eV. Searches at colliders  (LEP and LHC) set lower limits of order $10^{-3}$ eV (see e.g. \cite{Maltoni:2015twa}
and references therein).


 {\it Dark matter.}
Additional Weyl or Dirac fermions may constitute a component of the dark matter required by astrophysics and cosmology.
However typical cold dark matter candidates have masses above  $10^2-10^3$ eV. For a gravitino to be the dominant 
component of dark matter one needs $m_{3/2}\geq 90$ eV. So these additional Weyl fermions contributing to the 
Casimir energies do not seem to be natural candidates for dark matter.

Let us finally mention that ultralight fermionic states may contribute to the effective number of degrees of freedom $N_{eff}$ in cosmology. But the limits 
apply to particles who were at some point in thermal equilibrium with the SM and
decoupled before recombination. Details on bounds of dark radiation depend sensitively on
how and when the particle decoupled and hence need not apply to the light degrees of freedom here considered 
(see e.g. \cite{Brust:2013xpv,Pospelov:2010hj}).

Independently of any  motivation, it is clear that additional Weyl or Dirac fermions with masses relevant for the 
Casimir potential could be present in addition to the SM from e.g. hidden sectors or dark portals.  Here we will present results for the addition of one or two Weyl fermions.
The case of two Weyl fermions yields the same as the addition of one Dirac fermion or a gravitino.

\begin{figure}[t]
	\begin{center}
        \includegraphics[scale=0.33]{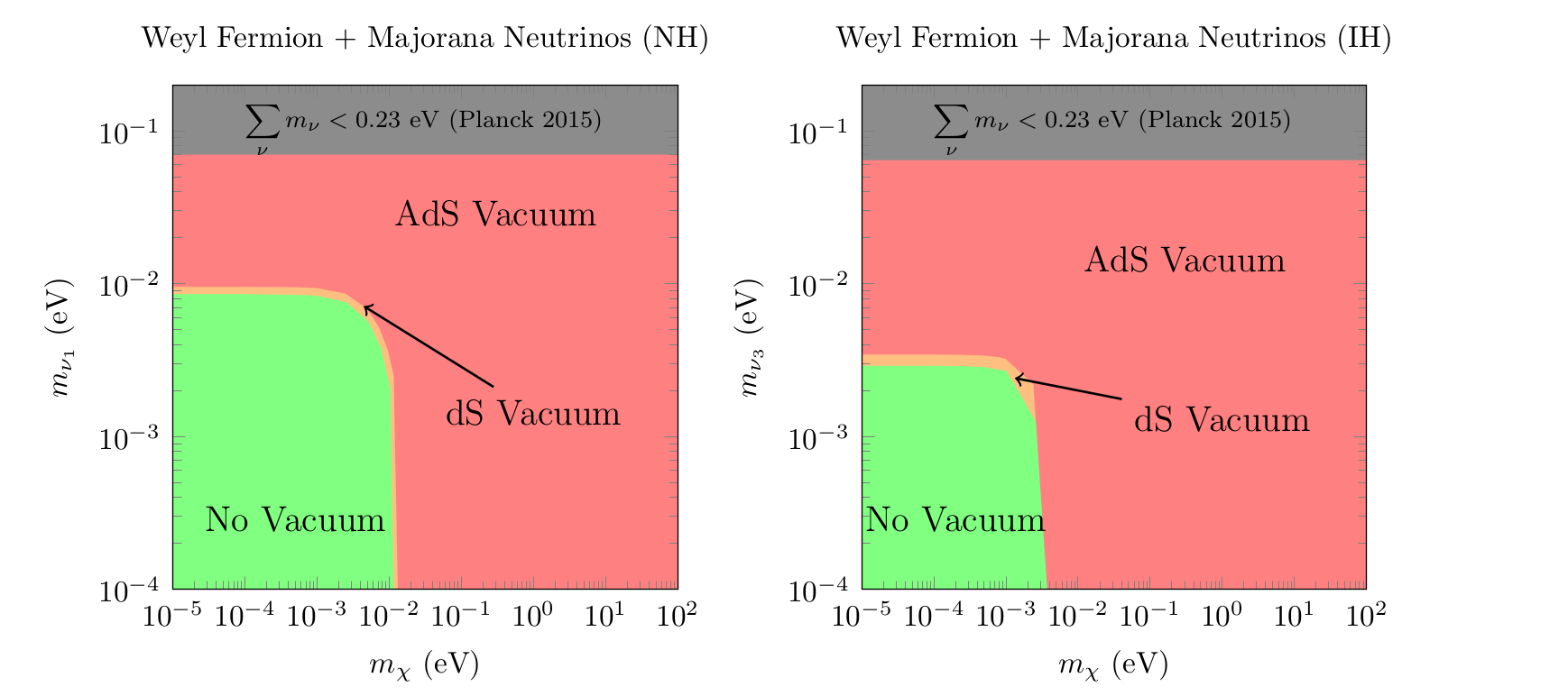}
		\caption{\footnotesize 3D vacua. Constraints on the lightest neutrino and Weyl fermion masses for the case of Majorana neutrinos, assuming no
		AdS 3D vacua forms. NH and IH stand for normal and inverted neutrino  hierarchies respectively.}
		\label{fig:majoranaaxino}
	\end{center}
\end{figure}
\begin{figure}[t]
	\begin{center}
        \includegraphics[scale=0.33]{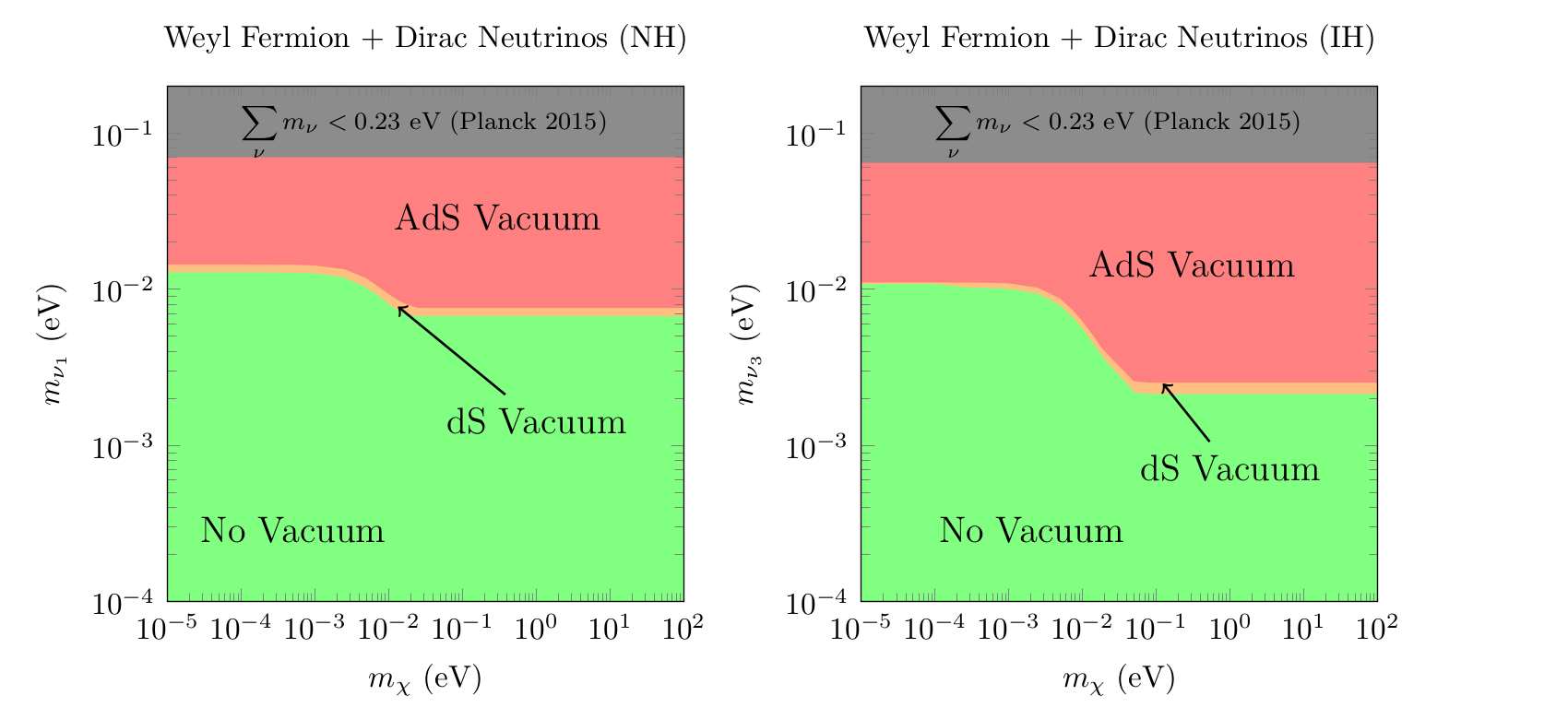}
		\caption{\footnotesize 3D vacua. Constraints on the lightest neutrino and Weyl fermion masses for the case of Dirac neutrinos, assuming no
		AdS 3D vacua forms. NH and IH stand for normal and inverted neutrino  hierarchies respectively}
		\label{fig:diracaxino}
	\end{center}
\end{figure}
\begin{figure}[t]
	\begin{center}
        \includegraphics[scale=0.33]{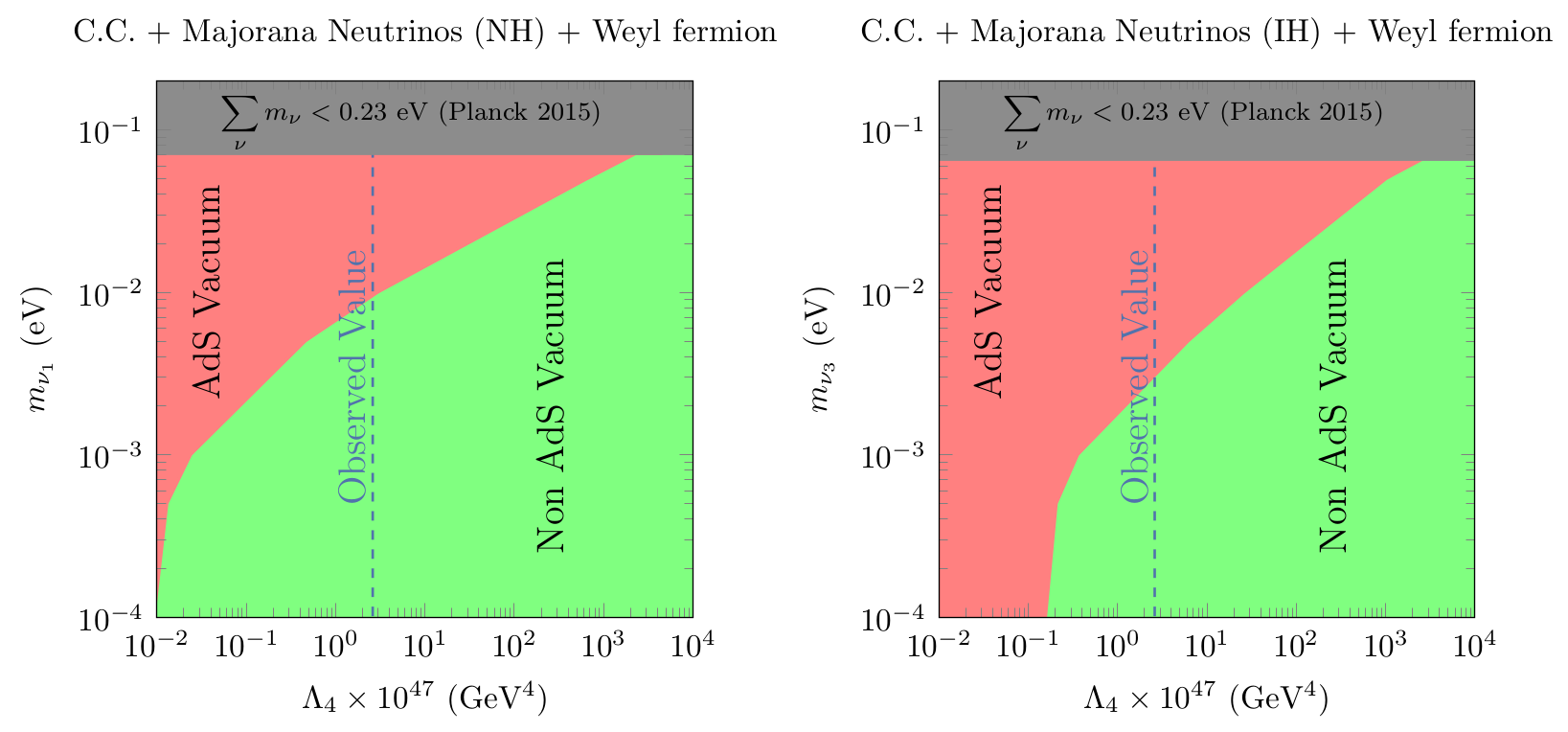}
		\caption{\footnotesize Majorana neutrinos with a Weyl fermion. Lower bound on the value of the 4D cosmological constant as a function of the
				lightest neutrino mass coming from absence of AdS 3D vacua when a Weyl fermion of mass $m_{\chi}=10^{-3}$ eV is added.  Left: NI. Right: IH.}
		\label{fig:ccmajoranaweyl}
	\end{center}
\end{figure}

\begin{figure}[t]
	\begin{center}
        \includegraphics[scale=0.33]{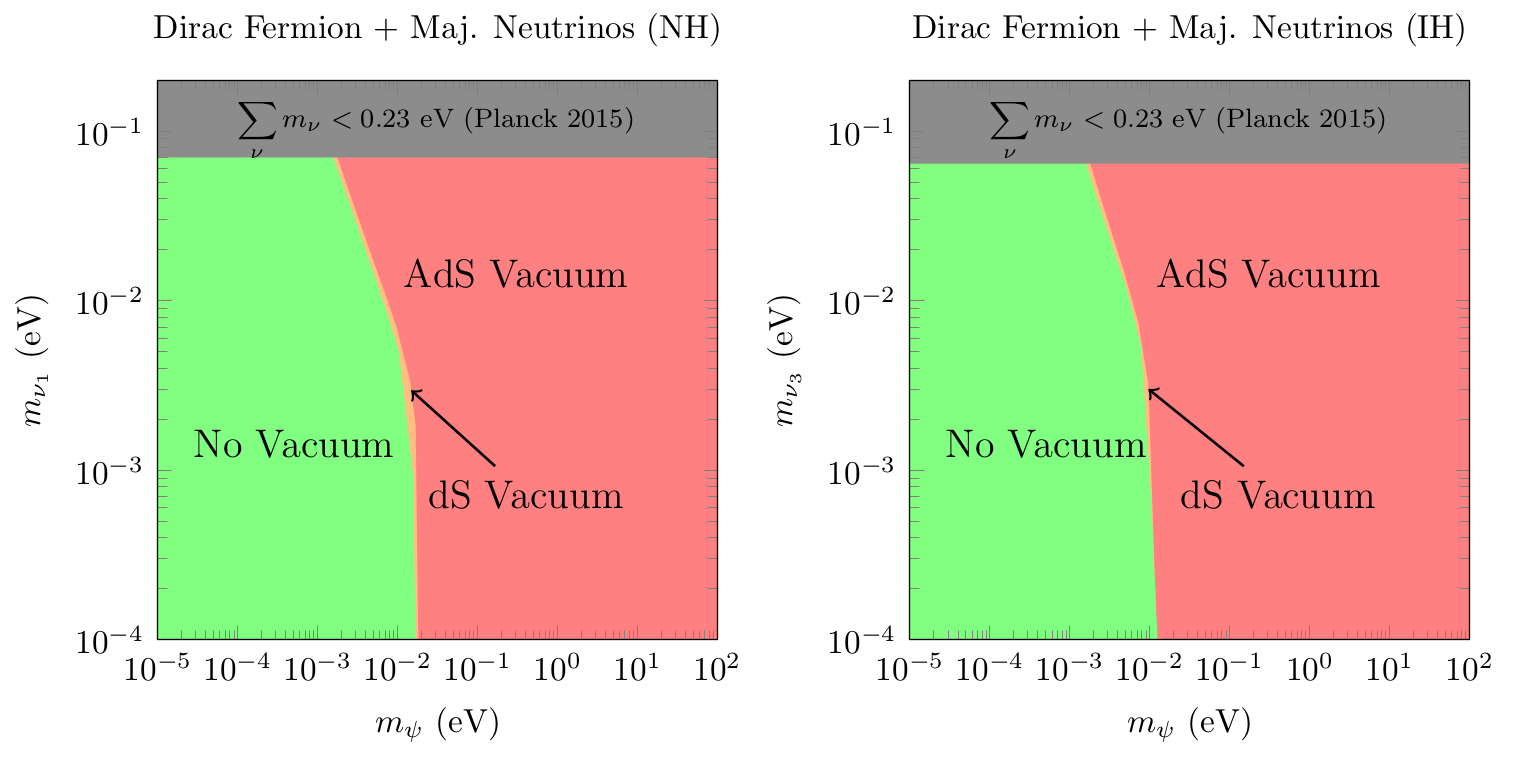}
		\caption{\footnotesize 3D vacua. Constraints on the lightest neutrino and Dirac/gravitino fermion masses for the case of Majorana neutrinos, assuming no
		AdS 3D vacua forms. NH and IH stand for normal and inverted neutrino  hierarchies respectively.}
		\label{fig:majoranagravi}
	\end{center}
\end{figure}
\begin{figure}[t]
	\begin{center}
        \includegraphics[scale=0.33]{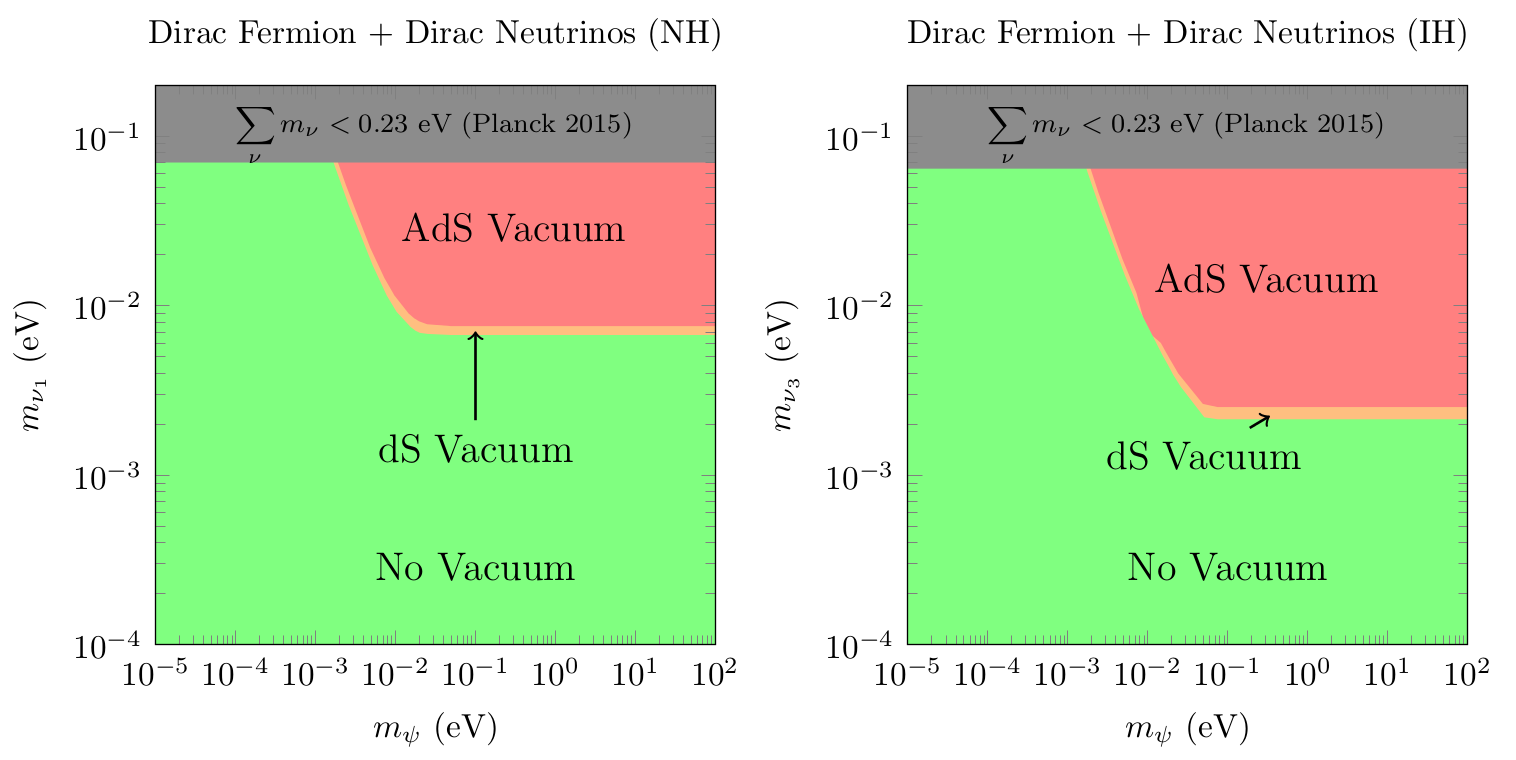}
		\caption{\footnotesize 3D vacua. Constraints on the lightest neutrino and Dirac  fermion masses for the case of Dirac neutrinos, assuming no
		AdS 3D vacua forms. NH and IH stand for normal and inverted neutrino  hierarchies respectively.}
		\label{fig:diracgravi}
	\end{center}
\end{figure}

\subsubsection{One Weyl fermion}

\begin{itemize}

\item {\it Majorana neutrinos.} The effect on the 3D Casimir vacua of the addition of one single Weyl fermion is shown in figs.\ref{fig:majoranaaxino}, both for the case
of NI and IH. Now the case of Majorana neutrinos becomes viable as long as the lightest neutrino is lighter than $m_{\nu_1}\leq 10^{-2}$ eV (NI) or
$m_{\nu_3}\leq 3\times 10^{-3}$ eV (IH).  The added Weyl fermion has also to be lighter than those values respectively.  Note that this values for the lightest 
neutrino Majorana masses would make complicated the detection in $\nu$-less double $\beta$-decay experiments in the case of normal hierarchy.
Planned experiments expect to reach values of order $10\times 10^{-3}$ eV for the effective Majorana mass in the double $\beta$ decay amplitude.
This would cover essentially all the IH problem in a model independent way. However, in the case of normal hierarchy only if the lightest neutrino es heavier 
than $10^{-2}$ eV the detection would be possible (see  e.g. \cite{Ostrovskiy:2016uyx}).

\item{\it Dirac neutrinos.} The effect on the 3D Casimir vacua of the addition of one single Weyl fermion is shown in figs.\ref{fig:diracaxino}, for both cases NH,IH. 
Recall that this case of Dirac neutrinos, unlike that of Majorana, was viable without the addition of any extra particle. We see that for Weyl fermion with mass
$m_\chi \geq 10^{-2}$ eV  one recover the limits of table \ref{tab:dirac}. For a lighter Weyl fermion the lower limit on the lightest neutrino mass become
slightly weaker, $m_{\nu_1}\leq 10^{-2}$  eV. Otherwise the vacua are not much altered.

\end{itemize}

\begin{figure}[t]
	\begin{center}
        \includegraphics[scale=0.33]{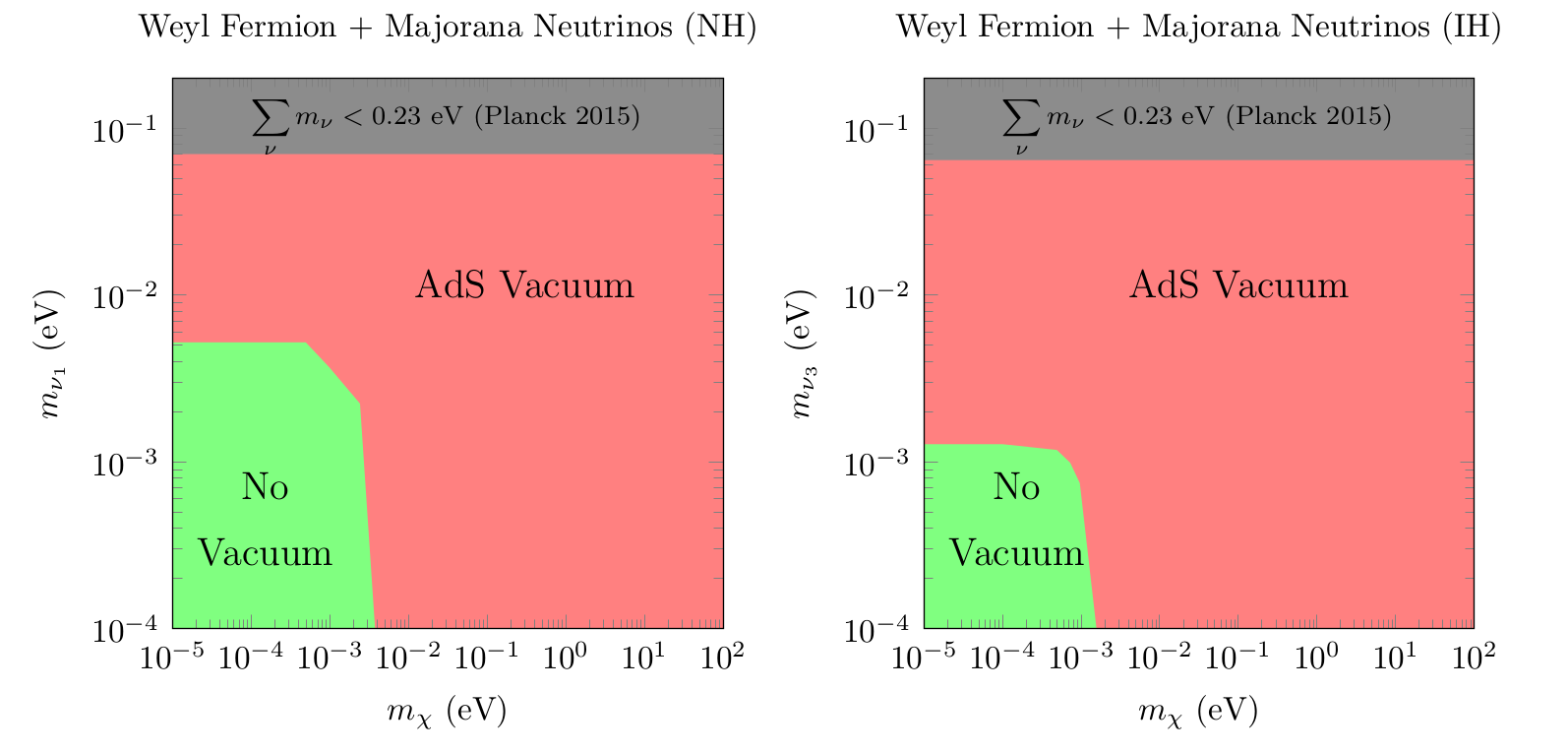}
		\caption{\footnotesize 2D vacua. Constraints on the lightest neutrino and Weyl fermion masses for the case of Majorana neutrinos, assuming no
		AdS 2D vacua forms. NH and IH stand for normal and inverted neutrino  hierarchies respectively.}
		\label{fig:majoranaaxino2d}
	\end{center}
\end{figure}

\begin{figure}[t]
	\begin{center}
        \includegraphics[scale=0.33]{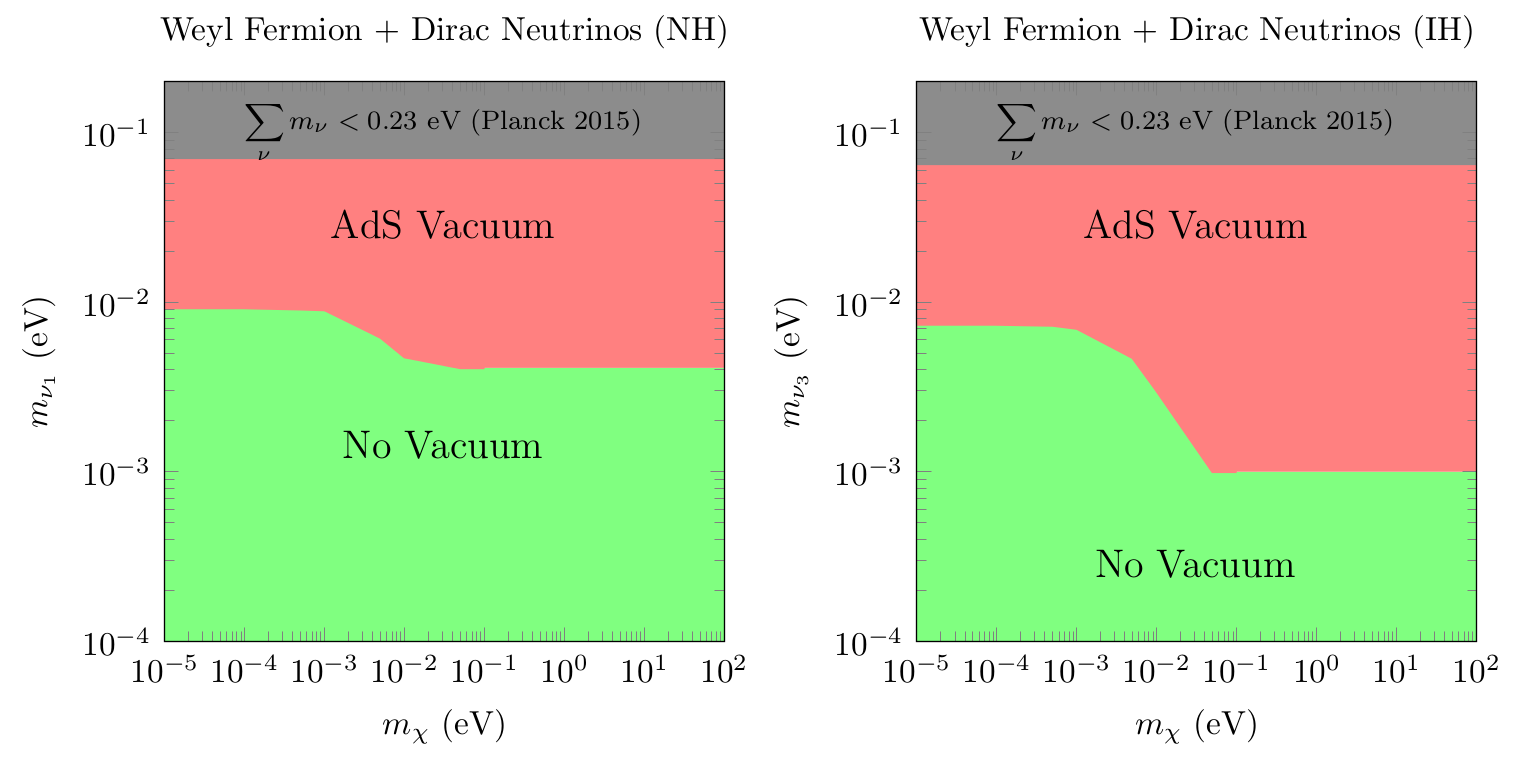}
		\caption{\footnotesize 2D vacua. Constraints on the lightest neutrino and Weyl fermion masses for the case of Dirac neutrinos, assuming no
		AdS 2D vacua forms. NH and IH stand for normal and inverted neutrino  hierarchies respectively}
		\label{fig:diracaxino2d}
	\end{center}
\end{figure}

We have also worked out the same study for the case of 2D vacua. The results are shown in figs.\ref{fig:majoranaaxino2d} and
\ref{fig:diracaxino2d}.  Compared to the case of 3D  the results are very similar though the obtained constraints are slightly stronger. Thus for the 
case of Majorana neutrinos, the lightest neutrino has to be lighter than $m_{\nu_1}\leq 5\times 10^{-3}$ eV for NI and even stronger 
$m_{\nu_3}\leq 10^{-3}$ eV for IH.

Given the fact that the addition of a Weyl fermion makes viable some regions of the scenario with Majorana neutrinos developing non AdS vacua, it is interesting to study if this is compatible with the observed value for the cc. As we studied in figs.~\ref{majcc} and \ref{fig:ccmajorananeutrino2d}, the current value of the cc is not compatible with the Majorana scenario, however this situation changes with the addition of a Weyl fermion. In fig.~\ref{fig:ccmajoranaweyl} the allowed values of the lightest neutrino versus the cc are depicted. We see that in contrast with figs.~\ref{majcc} and \ref{fig:ccmajorananeutrino2d} there are areas compatible with the cc value when the lightest neutrino mass is lighter than $m_{\nu_1}\leq 9\times 10^{-3}$ eV for NH and $m_{\nu_3}\leq 3\times 10^{-3}$ eV for IH for a mass of the Weyl fermion of $m_\chi=10^{-3}$ eV. These limits are dependent on the mass of the Weyl fermion since larger masses will reach the limit of figs.~\ref{majcc} and \ref{fig:ccmajorananeutrino2d}.

\subsubsection{One Dirac fermion/gravitino.}

\begin{itemize}

\item {\it Majorana neutrinos.} The effect on the 3D Casimir vacua of the addition of one  Dirac fermion is shown in figs.\ref{fig:majoranagravi}, both for the case
of NI andIH. Now the case of Majorana neutrinos is viable as long as the added Dirac fermion is sufficiently light, lighter than the two heaviest neutrinos.
Furthermore the upper bound on the mass of the lightest neutrino essentially disappears. This is important because it means that then a Majorana mass 
for the lightest neutrino could be detected in planned $\nu$-less double $\beta$-decay experiments  \cite{Ostrovskiy:2016uyx} also for the
NH case. 

\item{\it Dirac neutrinos.} The effect on the 3D Casimir vacua of the addition of one  Dirac fermion is shown in figs.\ref{fig:diracgravi}, both for the case
of NI andIH. The results are similar to those of an added Weyl fermion except for an important difference. As in the Majorana case, the upper bound on the mass 
of the lightest neutrino essentially dissappear if the added Dirac fermion has a mass smaller that $10^{-3}$ eV. 

\end{itemize}

\begin{figure}[t]
	\begin{center}
        \includegraphics[scale=0.33]{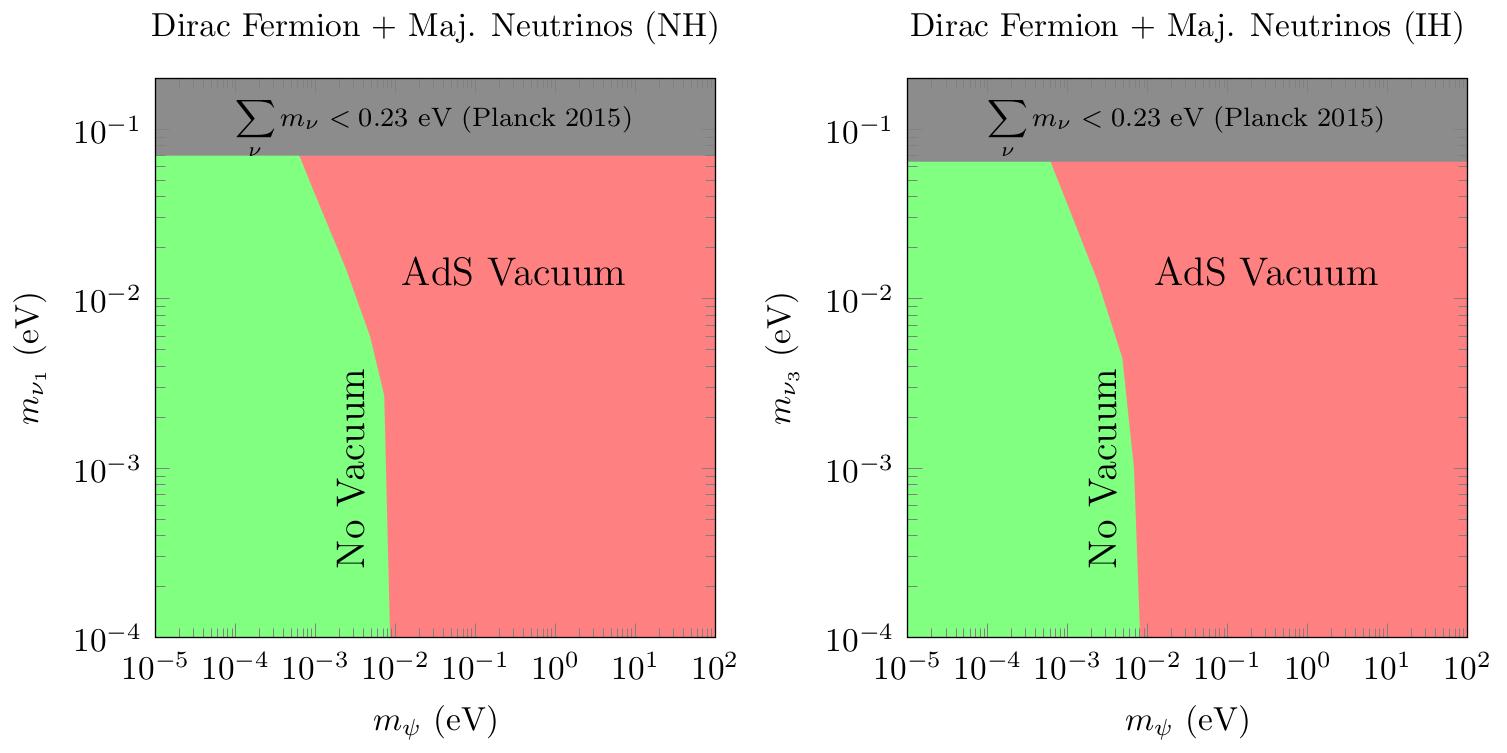}
		\caption{\footnotesize 2D vacua. Constraints on the lightest neutrino and Dirac/gravitino fermion masses for the case of Majorana neutrinos, assuming no
		AdS 2D vacua forms. NH and IH stand for normal and inverted neutrino  hierarchies respectively.}
		\label{fig:majoranagravi2d}
	\end{center}
\end{figure}

\begin{figure}[t]
	\begin{center}
        \includegraphics[scale=0.33]{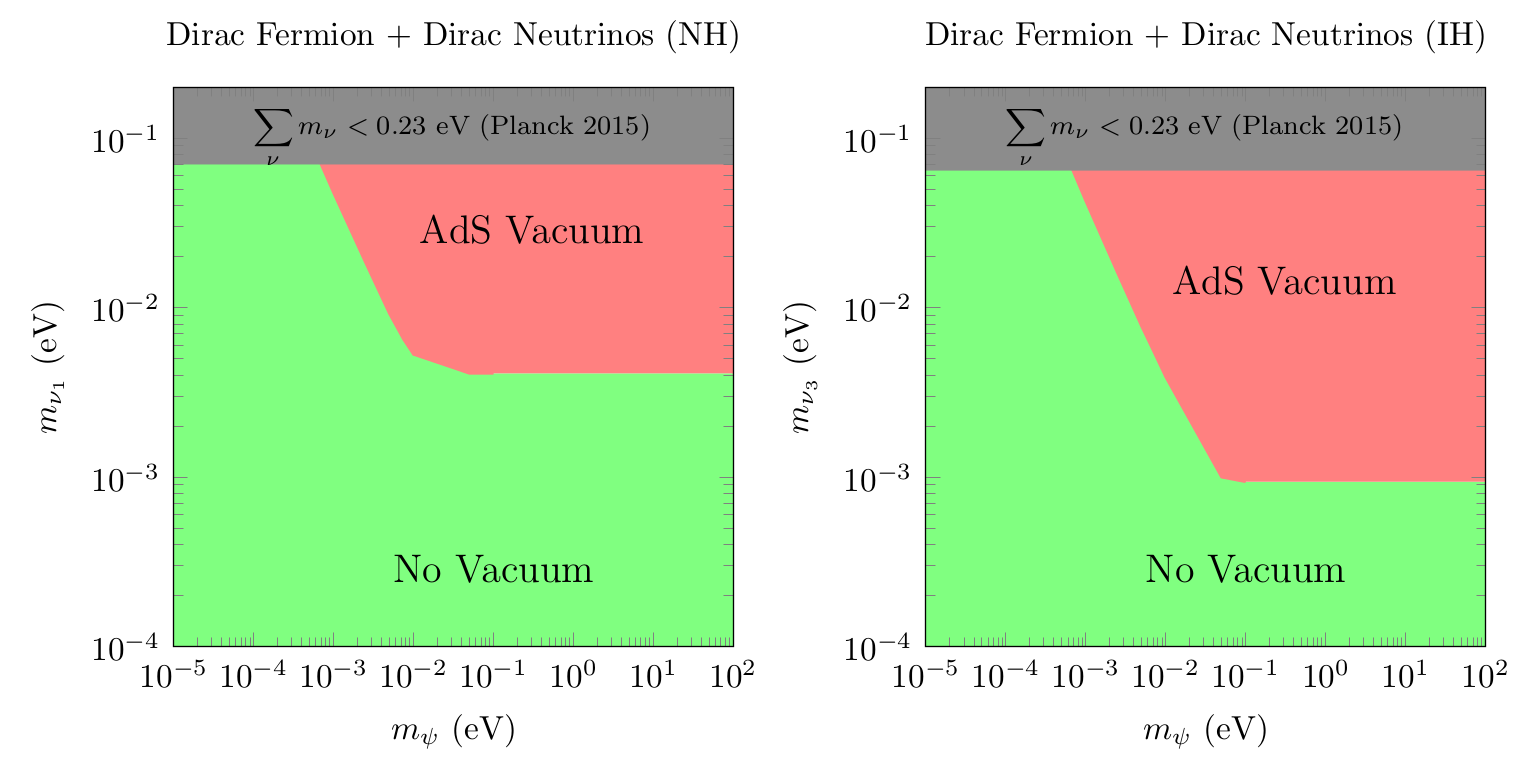}
		\caption{\footnotesize  2D vacua. Constraints on the lightest neutrino and Dirac  fermion masses for the case of Dirac neutrinos, assuming no
		AdS 2D vacua forms. NH and IH stand for normal and inverted neutrino  hierarchies respectively.}
		\label{fig:diracgravi2d}
	\end{center}
\end{figure}

We have also worked out the same study for the case of 2D vacua. The results are shown in figs.\ref{fig:majoranagravi2d} and
\ref{fig:diracgravi2d}.   They are almost identical to those we found for the 3D vacua. The limits are slightly stronger but 
one can barely note the difference.

As a summary, adding a Weyl or a Dirac fermion sufficiently light to Majorana neutrinos make the latter viable with the present constraints. 
The lightest Majorana neutrino would be amenable to planned $\nu$-less double $\beta$-decay if we add a Dirac fermion or a gravitino.



\subsection{Axions }

Axion-like particles  are natural candidates for BSM states populating the infrared sector of the SM.
Their shift symmetry $a\rightarrow a+2\pi f_a$  protects their masses from quantum corrections 
and make ultralight masses natural.  The best motivated such particle is the QCD axion which
is introduced to solve the strong CP problem of QCD. The mass of the QCD axion is given by 
(see e.g. \cite{villadoro} and references therein)
\begin{eqnarray}
m_a=\frac{z^{1/2}}{1+z}\frac{f_\pi m_\pi}{f_a},
\end{eqnarray}
where $f_\pi$ and $m_\pi$ are the pion decay constant and pion mass respectively and $z=m_u/m_d$. The mass of the axion can be written
\begin{eqnarray}
m_a=5.70\,{\rm eV}\,\frac{10^6\,{\rm GeV}}{f_a} .
\label{eq:qcdaxion}
\end{eqnarray}
Astrophysical and cosmological bounds for the  QCD axion constraint its decay constant to a range $10^8-10^{11}$ GeV,
so that the mass of the QCD axion lies in the range $m_a=(10^{-6}-10^{-2})$ eV, well in the ballpark of the neutrino mass scale, so that 
the standard QCD axion can significantly modify the   lower dimensional radion potential, as we describe below.

In addition to the QCD axion, the existence of other axion-like particles (ALP) has  been suggested for a variety of purposes.  For these 
ALP's the mass can vary in a very wide range.  A recently popular ALP is the {\it relaxion} \cite{relaxion}  in which the minimal model has a mass
as low as $m_a\simeq 10^{-25}$eV. In the formulation of relaxion in terms of 4-forms \cite{ourrelaxion}, the mass of the  relaxion is given by
$m_a=F_4/f_a$, where $F_4\simeq (10^{-3}eV)^2$ is the 4-form field strength.  An ALP  coupled to  4-forms (a {\it hierarxion} \cite{hierarxion}) and the 
Higgs particle has also been recently suggested in order to construct a landscape of values  for the Higgss mass. In this case the
ALP mass varies in a range $10^{-17}eV\ < m_a \ < 10^3eV$.  Axions  or ALP's may constitute the dark matter in the universe. Recently 
the case of ultralight scalars with mass $m_a\simeq 10^{-22}$ eV constituting what is called {\it fuzzy dark matter} has been 
studied  (see e.g.\cite{Hui:2016ltb}  and references therein).
All these  possible sources of axion-like particles could  if present contribute to the potential of the radion.

The axion contribution to the general effective potential would be negative due to its bosonic nature. In principle the axion contribution to the 3D potential reads,
\begin{eqnarray}
V_{a}=-\frac{r^3}{R^3}\frac{m_a^2}{4\pi^3 R}\sum_{n=1}^{\infty}\frac{1}{n^2}K_2(2\pi R m_a n).
\label{eq:axion}
\end{eqnarray}
However,  besides the Casimir contribution to the potential, there is an extra contribution to the potential from axionic fluxes \cite{ArkaniHamed:2007gg} .
The field strength of tha axion $(da)$ can be non-vanishing around the compact circle $S_1$ \footnote{Describing the axion in terms of its dual 2-form $B_2$, this flux
may be also understood as the flux of $H_3=dB_2$ through 3D space.},
\begin{eqnarray}
\oint_{S_1}da= f ,
\end{eqnarray}
with the flux $f$ quantized as $f=2\pi n f_a$, with $f_a$ the axion periodicity (decay constant). 
This flux contributes to the effective potential a piece
\begin{eqnarray}
V \propto \frac{f^2r^3}{4\pi R^4},
\label{eq:flux}
\end{eqnarray}
so that the full axion contribution to the potential is given by
\begin{eqnarray}
V_{a}^{tot}=\frac{f^2r^3}{4\pi R^4}-\frac{r^3}{R^3}\frac{m_a^2}{4\pi^3 R}\sum_{n=1}^{\infty}\frac{1}{n^2}K_2(2\pi R m_a n)  .
\label{eq:axiontotal}
\end{eqnarray}
Since the value of $f_a$ is enormously large in specific ALP's, the flux contribution completely overwhelms the Casimir contribution for non-vanishing fluxes
$n\not=0$. This destroys completely any possible Casimir induced vacua, and hence no constraint on low energy
parameters are obtained. 
 However, the conjecture tell us that there cannot exist any AdS  vacua and hence we have to 
study the possible vacua arising in the fluxless case $n=0$, which we analyze below. The effect of the axions 
on the 3D vacua depends on its number so we will distinguish two cases.

\subsubsection{One axion}

Let us consider first the case of one single axion
\footnote{Note  that the bounds  discussed in this section  hold as  well for  other light scalars, not necessarily axionic.}
. In the case of Majorana neutrinos the addition of an axion does not change things.  Since an axion contributes negatively,
an AdS vacuum still develops and becomes in fact deeper, since there are 6  fermionic degrees of freedom and 5 bosonic. 

\begin{figure}[t]
	\begin{center}
        \includegraphics[scale=0.33]{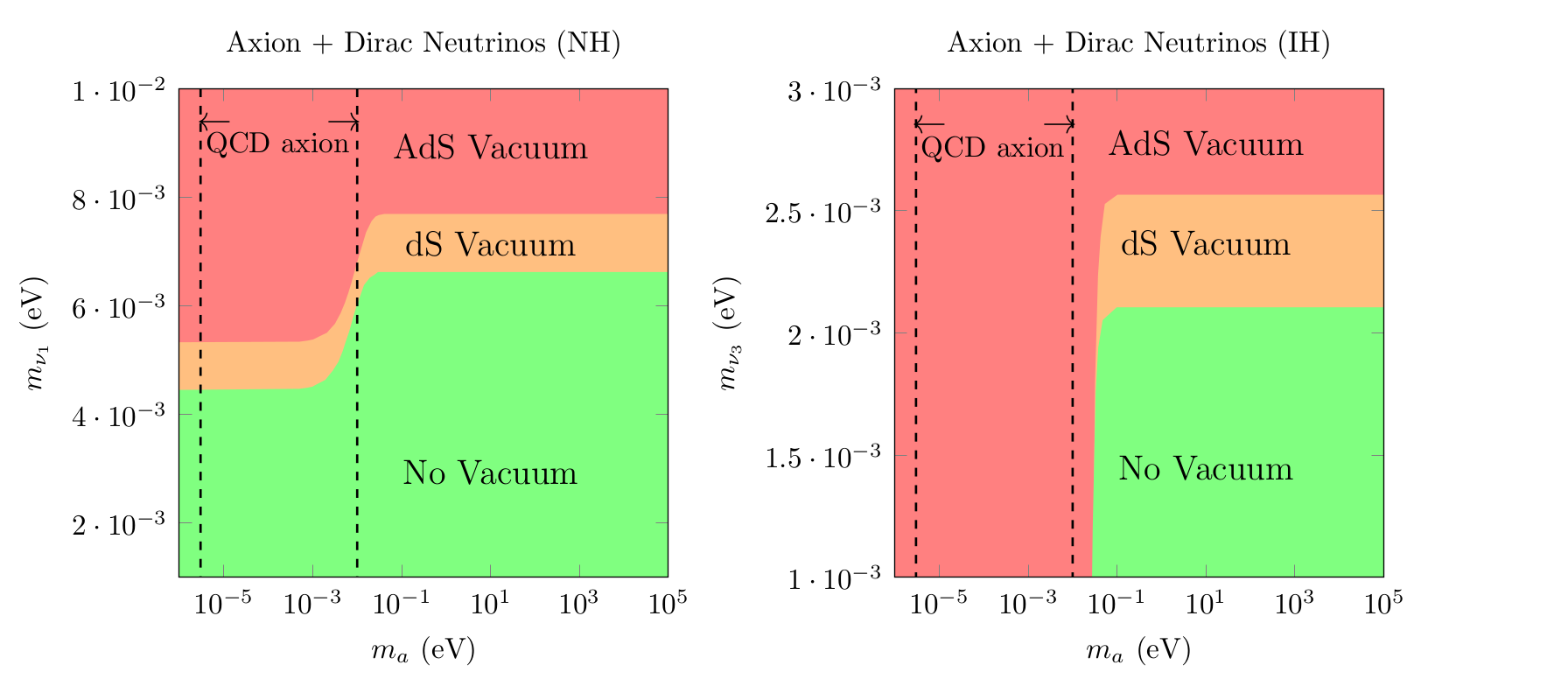}
		\caption{\footnotesize Contour plots for the appearence of different kind of 3D vacua in the mass of the lightest neutrino - mass of the axion plane. The range of the QCD axion is shown as an area delimited with dashed lines. On the left panel the case of NH is shown. For every mass of the axion we can find a bound on the lightest neutrino mass for which we can evade the AdS vacuum. On the right panel the case of IH is shown. In this case when the axion has a mass smaller than $m_a\leq 30$ meV, an AdS vacuum is  always formed.}
		\label{fig:neuaxion}
	\end{center}
\end{figure}

\begin{figure}[t]
	\begin{center}
        \includegraphics[scale=0.33]{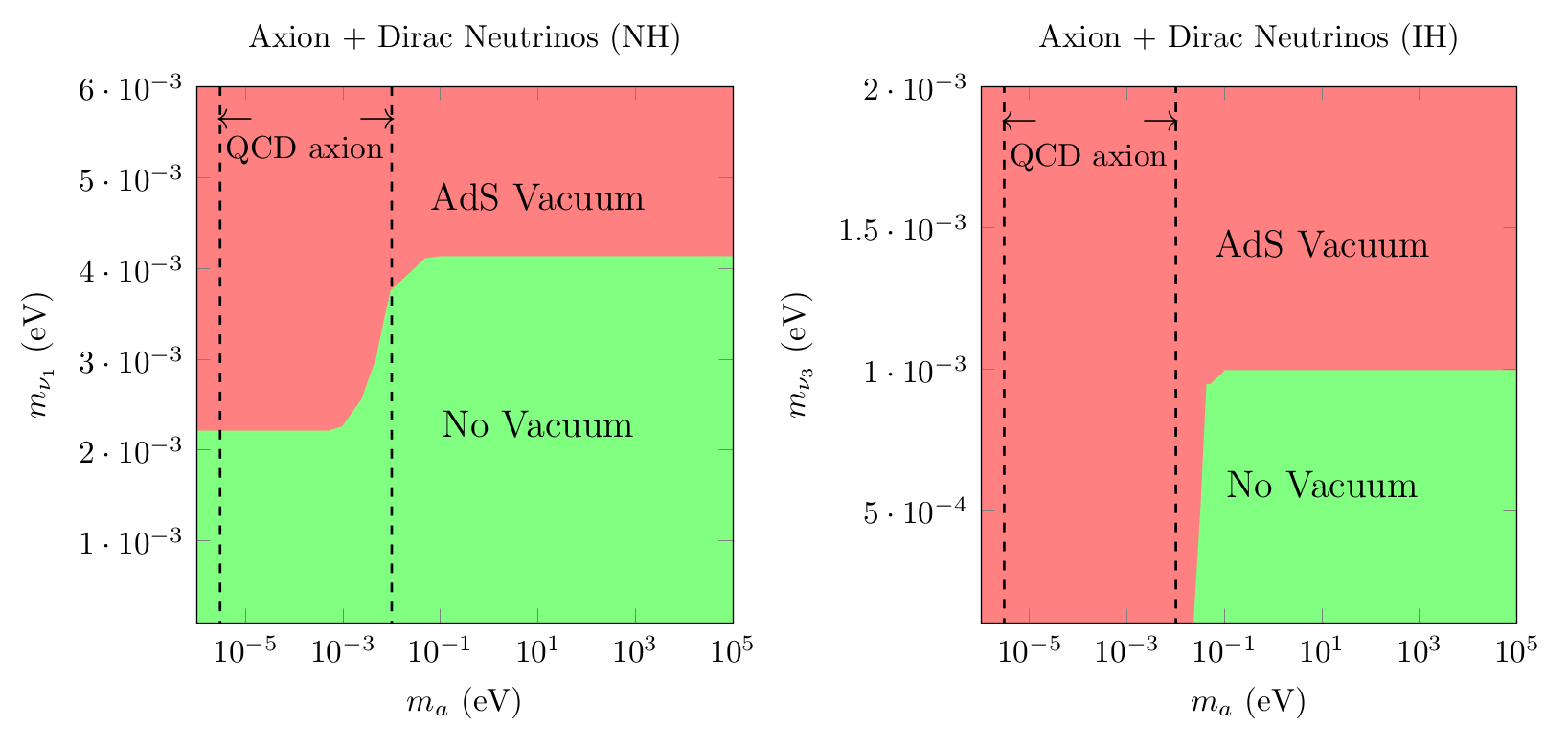}
		\caption{\footnotesize  Contour plots for the appearence of different kind of  2D vacua in the mass of the lightest neutrino - mass of the axion plane. The range of the QCD axion is shown as an area delimited with dashed lines. On the left panel the case of NH is shown. For every mass of the axion we can find a bound on the lightest neutrino mass for which we can evade the AdS vacuum. On the right panel the case of IH is shown.}
		\label{fig:neuaxion2d}
	\end{center}
\end{figure}

In the case of Dirac neutrinos, the negative contribution of the axion slightly modifies  some  of  the vacua, some of them could also change its nature or even create new vacua which
were absent in the axion-less case. The results depend on the relative magnitude of the axion mass and the mass of the heaviest neutrinos as well as whether the neutrino hierarchy 
is normal or inverted.
In Fig.~\ref{fig:neuaxion}  the effect of the axion for the different vacua formation is shown in the lightest neutrino mass and axion mass plane. For this plot we have assumed that $n=0$ so there is no contribution from the flux term. We have analysed masses of the axion from $10^{-6}$ eV to $10^{5}$ eV. We can see the  different effects that an axion could produce for NH and IH hierarchies:

\begin{itemize}

\item  {\it NH Dirac neutrinos}

In the NH case we see  that for axion masses above around $10^{-2}$ eV the number of light states becomes the same as in the axion-less case, and we recover the limit 
$m_{\nu_1}<7.7\times 10^{-3}$ eV. For axions lighter than $10^{-2}$ eV the effective number of degrees of freedom decreases one unit and the bound becomes 
stronger, $m_{\nu_1}< 5.35$ meV.

\item {IH Dirac neutrinos}

For IH, when we include an axion field an AdS vacuum is created even when the lightest neutrino mass is set to zero, $m_{\nu_3}=0$ eV. The reason for this behaviour is the fact that in IH there are two heavy states that even when the lightest neutrino mass is set to zero their masses are $m_{\nu_1}\sim m_{\nu_2}\sim 50$ meV. In this case there are 5 bosonic degrees of freedom against 4 fermionic ones below 50 meV, so an AdS vacuum is formed.  Note that the QCD region of axion masses would then be excluded for Dirac neutrino masses,
which is a strong result.
On the other hand, when the axion mass reaches the heavy neutrino states masses the fermionic degrees of freedom start contributing to the effective potential. In that sense, when the mass of the lightest neutrino is set to zero, $m_{\nu_3}=0$ eV, one finds that for masses of the axion greater than $m_a>24.8$ meV the AdS vacuum becomes a dS one. For instance, for an axion mass of $m_a=50$ meV or larger the limit of the lightest neutrino mass in order to avoid an AdS vacuum is $m_{\nu_3}=2.5$ meV. 

\end{itemize}
A summary of the constraints for axion and lightest Dirac neutrino masses is shown in table (\ref{tab:axion}). Very similar results and constraints are 
obtained in the case of a compactification down to 2D as can be seen from  fig.(\ref{fig:neuaxion2d}).

\begin{table}
\begin{center}
\begin{tabular}{|c | c | c |}
$m_a$ & NH & IH \\
 \hline
$\lesssim 10^{-4}$ eV & $m_{\nu_1}> 5.35$ meV & $m_{\nu_3}>0.0$ meV\\
$10^{-3}$ eV & $m_{\nu_1}> 5.4$ meV& $m_{\nu_3}>0.0$ meV\\
$10^{-2}$ eV & $m_{\nu_1}> 6.87$ meV & $m_{\nu_3}>0.0$ meV\\
$\gtrsim 10^{-1}$ eV & $m_{\nu_1}> 7.7$ meV & $m_{\nu_3}>2.55$ meV\\
\hline
\end{tabular}
\caption{Upper bound on the lightest neutrino mass for NH and IH up to which an AdS vacuum is formed for different QCD axion masses.}
\label{tab:axion}
\end{center}
\end{table}

\subsubsection{Multiple axions and axiverse}

For more than one axion-like particle the situation may change in an important way.
The reason is that if a sufficiently large number of axions have their masses in the neutrino mass range or below, they may make unstable 
any of the Casimir vacua here discussed. The reason is that they give rise to a negative contribution to the potential which may dominate 
it if the total number of bosonic  degrees of freedom exceeds the number of fermionic ones
\footnote{Note that  the discussion here also applies to the case of  very light but massive gauge bosons, which would contribute as three scalars.}.
Again, the number of axions required to destabilize the AdS vacua will depend on the Majorana or Dirac nature of neutrinos:

\begin{itemize}

\item {\it Majorana neutrinos plus multiaxions} .
If there are  $\geq 2$ light axions, the number of fermionic degrees of freedom is smaller than the number of bosonic ones, and AdS vacua do not form.
So in principle this is a simple way in which Majorana neutrino masses can be made compatible with the absence of dangerous AdS vacua.
The situation may be in fact slightly more complex. Indeed, as $R$ decreases, other particle thresholds become eventually relevant
\cite{ArkaniHamed:2007gg}. Up to the QCD scale we have
the electron, then the muon, the pion, the kaons and the $\eta$.  Just above the electron threshold there are 10 fermionic degrees of freedom, so we would need
6 or more axions for such local minima not to develop. And above the muon threshold there are 14 fermionic degrees of freedom so 10 or more
axions are needed to avoid some local AdS minimum. Higher thresholds involves bosons ( at least up to the QCD scale).  So we may need more than 10 
light axions to make the AdS vacua not to form.  Still these extra AdS vacua involving higher thresholds  may be unstable to decay into other vacua for larger 
$R$, so that perhaps 2 axions may be enough to avoid stable AdS vacua.

\item{\it Dirac neutrinos plus multiaxions}. 
In this case AdS vacua may be avoided already in the absence of axions. Now if we have more than $8=12-4$ axions any neutrino 
Casimir AdS vacua becomes unstable, so that no constraints on neutrino masses is obtained. To avoid formation of additional AdS at the 
muon threshold we would need a total  $8+4+4=16$ axions. But again this may be too constraining if these additional vacua
are not stable.

\end{itemize}

In summary, a simple way to avoid unwanted  Casimir AdS vacua both for Majorana and Dirac neutrinos  is to have
multiple axions (and/or gauge bosons)  in the ultralight spectrum of the theory. This would fit  well with the idea of an Axiverse, as suggested
in ref.\cite{axiverse}.

\section{The Electro-Weak  hierarchy problem and the cosmological constant}

The essential ingredient  to minimally avoiding  3D, 2D  AdS vacua  is having 4 fermionic degrees of freedom sufficiently light (lighter than $\simeq \Lambda_4^{1/4}$) so
as to cancel the negative contribution coming from the photon and graviton,  before the radion potential becomes negative,  as the compact radii decrease.   It is then clear that,
{\it for a fixed value of $\Lambda_4$},  the mass of these lightest fermionic degrees of freedom is bounded from above as it is clearly shown in figs.~\ref{majcc} and~\ref{dircc}.
In the case of Majorana neutrinos, in addition to the ligthest neutrino, an additional Weyl fermion state lighter than $10^{-3}$ eV must also be added if we want
to avoid AdS vacua. But again
one observes in fig.~\ref{fig:ccmajoranaweyl} that there is an upper bound on the mass of the lightest neutrino (both in normal NI and IH).
Similar results are obtained in compactifications to 2D.

\begin{figure}[t]
	\begin{center}
        \includegraphics[scale=0.33]{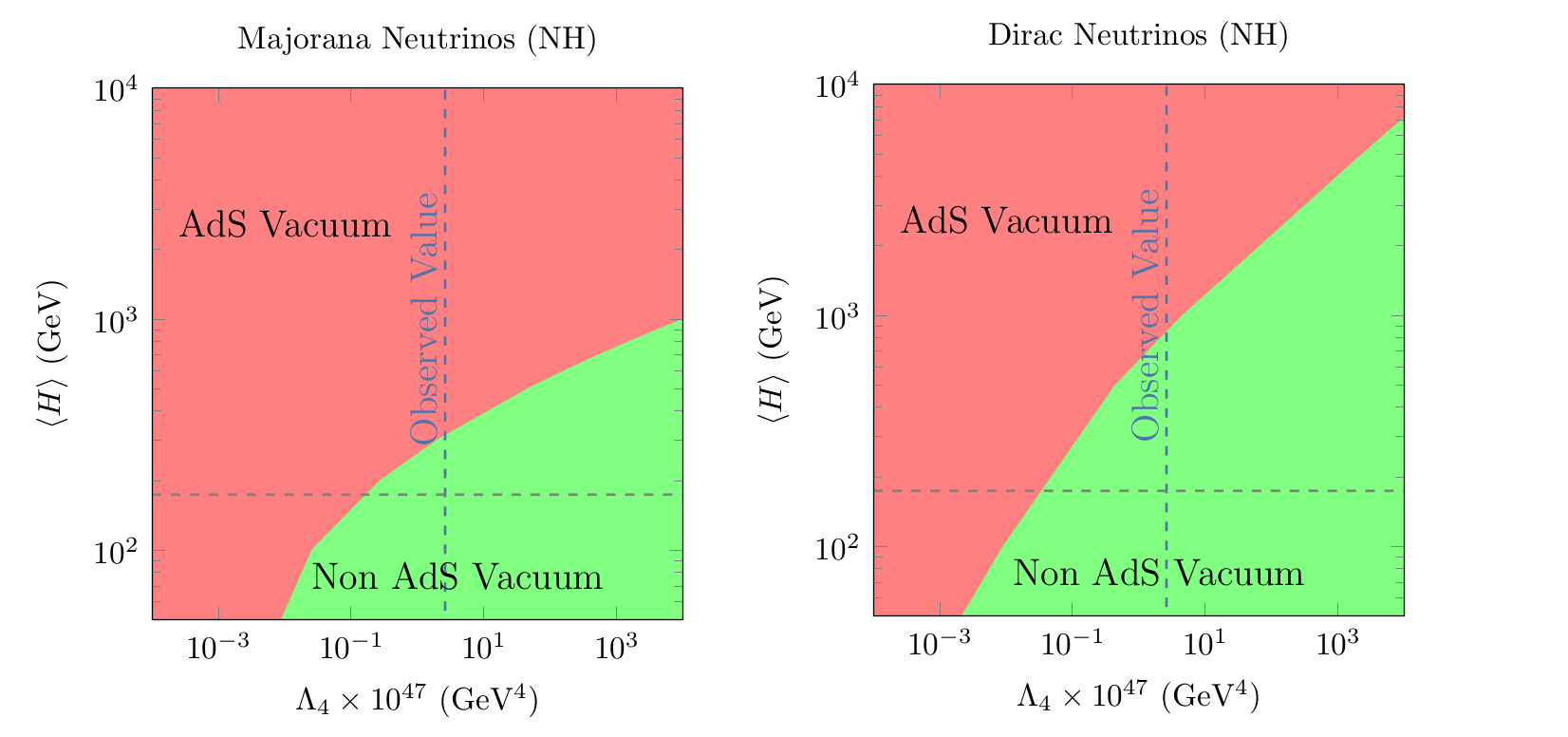}
		\caption{\footnotesize  Constraints on the EW scale and the cosmological constant. On the left panel the case of Majorana neutrinos and normal hierarchy is shown, in the presence of an additional Weyl fermion of mass $m_\chi=10^{-3}$ eV.  We have assumed $Y=10^{-3}$ and $M=10^{10}$ GeV. On the right panel the case of Dirac neutrinos and normal hierarchy assuming a Yukawa coupling $Y=10^{-14}$ is depicted. }
		\label{fig:vevcc}
	\end{center}
\end{figure}

If neutrinos are Majorana one sees from table~\ref{tab3D2D} that $m_{\nu_1}\lesssim  5 (1)\times 10^{-3}$ eV$\sim  2 (0.4)\times \Lambda_4^{1/4}$
for NI (IH) respectively. If the lightest neutrino Majorana mass is induced from a standard see-saw mechanism one obtains (e.g for NI)
\footnote{Of course, one only obtains a useful bound if the lightest neutrino has non-zero mass.}
\beq
\frac {(Y_{\nu_1}<H>)^2}{M} \ \lesssim\ 2\times \Lambda_4^{1/4} \ \longrightarrow \  \\
<H> \ \lesssim  \ \frac {\sqrt{2}}{Y_{\nu_1}} \sqrt{M \Lambda_4^{1/4}}  \ .
\label{eq:hierarchy}
\eeq 
where $M$ is the scale of lepton number violation in the see-saw.
Thus one gets the interesting conclusion that, for a  given fixed  c.c. $\Lambda_4$ and fixed Yukawa coupling, the EW scale  is bounded above by
the geometric mean of the cosmological constant scale and the lepton number violation scale $M$. Thus, e.g. for $Y_{\nu_1}\simeq 10^{-3}$ and
$M\simeq 10^{10}-10^{14}$ GeV, one gets $<H> \lesssim  10^2-10^4$ GeV.  Larger EW scales would yield (for fixed Yukawa) too large lightest neutrino mass and
AdS vacua would be generated. In other words, consistency with quantum gravity requires that a very small 4D cosmological constant should come accompanied by a big hierarchy between the EW scale and $M$. 
On the left panel of figure \ref{fig:vevcc}   we depict the constraints on the EW scale (parametrised by the Higgs vev) and the 4D cosmological constant for fixed $Y=10^{-3}$ and $M=10^{10}$ GeV, leading to the aforementioned upper bound on the EW scale.  To obtain this figure we have used the bounds provided by  fig.~\ref{fig:ccmajoranaweyl}.
Similar results apply for the case of inverted neutrino mass hierarchy and 2D vacua.

In the case of Dirac neutrinos one rather gets 
\beq
 <H> \  \lesssim  \  1.6(0.4) \frac {\Lambda_4^{1/4} }{Y_{\nu_1}}
 \eeq
 for NI(IH). 
  Now,  for fixed Yukawa coupling the EW scale is again bounded above by the 4D cosmological constant. In the Dirac case, 
though, the Yukawa
coupling needs to be extremely small to match the scale of observed neutrino masses
\footnote{In the case of Dirac neutrinos, one can also apply the argument in the opposite direction, 
 to explain why at least one of the neutrinos has a Yukawa $\lesssim 10^{-14}$.
Indeed {\it for fixed $\Lambda_4$ and EW scale}, one lightest neutrino  with a Yukawa coupling $\lesssim 10^{-14}$ would be enough  to
avoid the existence of 3D, 2D AdS vacua. However the other two neutrino generations would not be constrained from such arguments.}. But again, the smallness of the cosmological constant implies in turn a small EW scale in order to be consistent with quantum gravity. This relation is shown on the right panel of figure \ref{fig:vevcc} for fixed Yukawa coupling $Y=10^{-14}$.

Note that from the point of view a low energy  Wilsonian field theorist the smallness of the EW scale is surprising because there is apparently nothing preventing 
the Higgs mass to grow up to the UV cut-off scale. That is the hierarchy problem. If that  huge UV mass squared is negative,  that would give rise to EW breaking close to the UV scale. 
We now see that,  from the WGC point of view  here considered,  that situation would not be possible (for fixed $\Lambda_4$) because AdS vacua would then be generated
at 3D and 2D compactifications. The smallness of the EW scale becomes, therefore, indirectly related to the smallness of the cosmological constant. 

The other option is having a positive UV scale mass for the Higgs, i.e., no Higgs at low energies at all.  That situation 
turns out to be also inconsistent with the WGC.  Indeed, starting with the SM with just fermions, gauge bosons and no Higgs, the theory has a global
accidental $U(6)_R\times U(6)_L$ symmetry in the quark sector. Once QCD condensation takes place, the symmetry is broken to the diagonal $U(6)$ and
36 Goldstone bosons appear. Out of those 3 are swallowed by the $W^\pm$ and $Z$ bosons. These large number of bosons outnumbers the 
massless leptonic  degrees of freedom  which are 18 or 24 if neutrinos are Dirac. This makes again an AdS vacuum to develop. In summary, for a SM with fixed Yukawa couplings and the observed c.c., having a light Higgs is 
arguably the only way to scape inconsistency with quantum gravity.

One can convert the above bound  on the EW scale into a prediction if one assumes that indeed this scale is fully fixed 
by this constraint.  By  this we mean that any slight increase on the Higgs vev would put the theory into the swampland, so that the bounds are 
saturated.
If this is the case and the WGC provides the full explanation for the EW gauge hierarchy, the mass of the lightest neutrino
should be at the value given by its upper bound. Then the predicted lightest neutrino masses are shown in table \ref{tablapred}.

\begin{table}[htb] \footnotesize
\renewcommand{\arraystretch}{1.25}
\begin{center}
\begin{tabular}{|c||c|c|}
\hline      
        &      NI    &   IH   \\
\hline
\hline
 Dirac &    $ m_{\nu_1}=4.12\times 10^{-3} $ \ eV &     $m_{\nu_3}= 1.0\times 10^{-3}$ \ eV  \\
 \hline 
 Majorana &   $m_{\nu_1}=5\times 10^{-3}$  \ eV &      $m_{\nu_3}=1.0\times 10^{-3} $\ eV \\
 \hline 
\end{tabular}
\end{center} \caption{ Predictions for the mass of the lightest neutrino assuming the WGC constraint is saturated and the EW hierarchy 
is thus explained by the bounds coming  absence of 2D(3D)  AdS vacua. The Majorana case assumes the existence of an additional 
Weyl fermion with mass lighter than $4(2)\times 10^{-3}$ eV (otherwise the Majorana case is already ruled out).
}
\label{tablapred} 
\end{table}
Thus e.g., if  we were able to measure the mass of the electron neutrino at a $\nu$-less double
beta decay experiment,  and it was established that  the neutrinos have NH(IH), a mass
 $m_{\nu_1}=5(1)\times 10^{-3}$ eV would be a strong indication that the origin of the EW hierarchy lies in the above WGC arguments.

Note in closing that the above WGC arguments not necessarily imply the absence of any new physics above the EW scale like e.g. low energy SUSY.
 The latter could be present for other purposes like dark matter and in particular the stability of
the Higgs potential at high energies. In fact the WGC arguments could perhaps explain  the existence of a little hierarchy problem in the MSSM or the fine-tuning  in models like split SUSY or high scale SUSY.

\section{Discussion}

In the present paper we have rexamined the implications of
 the recent conjecture of Ooguri and Vafa suggesting  that theories with consistent
quantum gravity cannot have  AdS 
stable non-SUSY vacua.  When applied to the vacua obtained from compactifications of the SM to  3D and 2D studied by Arkani-Hamed et al. 
one obtains strong constraints on neutrino masses and other possible BSM very light particles in terms of the c.c.. Furthermore
one also obtains a new understanding of the EW hierarchy problem  from consistency with
quantum gravity.

As we have emphasized, a crucial point to obtain such constraints is the issue of the stability of these AdS vacua.  Although a decay of these 
SM vacua to a  Witten {\it bubble of nothing } is ruled out due to the periodic boundary conditions of the fermions, it is more difficult to exclude other
sources of instability. In particular, if the SM is embedded into a landscape of vacua as suggested in string theory, the  4D vacuum transitions which
should occur to populate the vacua would have a reflection on lower dimensions, giving rise to instabilities. However, as we discussed in section 2,
our knowledge of the structure of the landscape of vacua in string theory is far from complete and one could envisage a situation in which
the barriers around the SM are huge,  and it could be that the lower dimensional Casimir vacua here discussed were stable.

In spite of the uncertainties concerning vacuum stability, we think it is interesting to work out in detail what would be
the consequences if indeed the Ooguri Vafa conjecture is correct and the Casimir AdS SM vacua were indeed stable.
It turns out that this assumption leads to quite interesting physical constraints for the 4D cosmological constant, the masses of neutrinos,
 extra additional light particles BSM and even the possible origing of the EW hierarchy.

One first interesting result is the existence of a lower bound on the value of the c.c. in terms of the light degrees of freedom 
of the SM Casimir potential. One can obtain an approximate analytic expression of the form
\beq
\Lambda_ 4 \ \geq \  A(\sum_im_i^2)^2 \ -\ B\sum_im_i^4 \ .
\eeq
This is interesting because, as far as we are aware, this is the only known suggestion for a non-vanishing 
value of the c.c. related to neutrino masses and independent of any cosmological argument (dark energy). 
Before evidence for an accelerating universe was found, it was widely believed that $\Lambda_4=0$. 
The conjecture here studied would have implied the existence of a 4D c.c. to avoid inconsistent AdS vacua,
independently of any cosmological argument.

We find that the existence  or not of dangerous lower dimensional SM AdS vacua is very sensitive both to the
value of $\Lambda_4$, neutrino masses and possible BSM extensions. We have done a systematic study 
of this dependence for both 3D and 2D SM vacua and the summary is shown in table \ref{tab3D2D}.
The results for 2D and 3D vacua are quite similar, although bounds coming from the absence of
2D vacua are slightly stronger.

\begin{table}[htb] \footnotesize
\renewcommand{\arraystretch}{1.25}
\begin{center}
\begin{tabular}{|c||c|c||c|c|}
\hline  Model &
                     Majorana \ (NI) &   Majorana \ (IH) &   Dirac \ (NH)   &   Dirac \ (IH)  \\
\hline
 SM (3D) &   no  &     no  &     $m_{\nu_1}\leq 7.7\times 10^{-3}$      &   $m_{\nu_3}\leq  2.56\times 10^{-3}$  \\
 \hline 
 SM(2D) &   no  &     no  &     $m_{\nu_1}\leq 4.12 \times 10^{-3}$      &   $m_{\nu_3}\leq  1.0\times 10^{-3}$  \\
 \hline 
 SM+Weyl(3D)      &    $m_{\nu_1}\leq 0.9 \times 10^{-2}$    &    $m_{\nu_3}\leq  3 \times 10^{-3}$  & $m_{\nu_1}\leq 1.5\times  10^{-2}$  &   $m_{\nu_3}\leq 1.2\times  10^{-2}$  \\
                         &     $m_f \leq 1.2 \times 10^{-2}$   &                  $m_f \leq  4\times 10^{-3}$ &     & \\
                         \hline 
 SM+Weyl(2D)      &    $m_{\nu_1}\leq 0.5 \times 10^{-2}$    &    $m_{\nu_3}\leq  1 \times 10^{-3}$  & $m_{\nu_1}\leq 0.9 \times  10^{-2}$  &   $m_{\nu_3}\leq 0.7 \times  10^{-2}$  \\
                         &     $m_f \leq 0.4 \times 10^{-2}$   &                  $m_f \leq  2\times 10^{-3}$ &     & \\
\hline
 SM+Dirac(3D)    &      $m_f\leq   2\times 10^{-2}$  &    $m_f \leq  1 \times 10^{-2}$  &   yes     &     yes    \\
 \hline
 SM+Dirac(2D)   &      $m_f\leq   0.9\times 10^{-2}$  &    $m_f \leq  0.9 \times 10^{-2}$  &   yes     &     yes    \\
 \hline
 SM+1\ axion(3D) &   no  &  no  &      $m_{\nu_1}\leq 7.7\times 10^{-3}$   &  $m_{\nu_3}\leq 2.5\times 10^{-3}  $ \\
                       &         &        &                                                                    &  $m_a\geq 5\times 10^{-2}$ \\
 \hline
 SM+1\ axion(2D)  &   no  &  no  &      $m_{\nu_1}\leq 4.0 \times 10^{-3}$   &  $m_{\nu_3}\leq 1 \times 10^{-3}$   \\
                       &         &        &                                                                    &  $m_a\geq 2\times 10^{-2}$ \\
                       \hline 
 $\geq 2(10)$\ axions  &  yes &  yes & yes & yes\\
\hline
\end{tabular}
\end{center} \caption{ Conditions on neutrino, fermion and axion  masses (in eV)  from the absence of  3D and 2D SM vacua. Here {\it yes}  means that no AdS value forms independently of the values of parameters, {\it no} means the opposite. Note that the 2D constraints are slightly stronger than the 3D constraints but follow a similar patern.
Majorana neutrino masses accessible to  $\nu$-less double $\beta$-decay require the existence of either at least 2 additional 
weyl  spinors or 2 or more scalars (e.g. axions or ultralight vector bosons).}
\label{tab3D2D} 
\end{table}

Perhaps the most attractive setting for neutrino masses is that of Majorana neutrinos (from a see-saw mechanism) in normal hierarchy.
If no additional BSM states are around, Majorana neutrinos are not consistent with the bounds from absence of AdS vacua here discussed, as already pointed out in
\cite{OV}. However we have found that slight modifications like the addition of a  Weyl fermion  with $m_\chi\leq 4\times 10^{-3}$ eV
is  sufficient to ensure the absence of dangerous vacua. This requires a lightest neutrino mass $m_{\nu_1}\leq 5\times 10^{-3}$ eV, difficult to measure
in planned $\nu$-less double $\beta$-decay experiments, if the hierarchy is normal. However if there are 2 light Weyl spinors (or a Dirac fermion or gravitino) this upper bound on 
the lightest neutrino mass dissappears and Majorana masses may be detectable at those experiments. 
 This is also the case if in addition of the SM we have a sufficiently large 
number of light bosonic states making any would be AdS vacua to dissappear, for any value of neutrino masses.
These may come from  a multiple set of axions or ultralight vector bosons.

If the neutrinos are Dirac,  dangerous AdS vacua may be avoided even in the absence of new physics as long as the
lightest neutrino  has $m_{\nu_1}\leq 4.1\times 10^{-3}$ eV  for NH ($m_{\nu_3}\leq 1\times 10^{-3}$ eV for IH). If one Weyl
fermions are added these bounds are increased to  $m_{\nu_1}\leq 0.9\times 10^{-2}$ eV for NI ($m_{\nu_3}\leq 0.7\times 10^{-2}$ eV for IH).
If instead we add a light Dirac/gravitino state, bounds on the lightest neutrino mass dissappear and dangerous AdS vacua are altogether avoided.

An interesting light addition to the SM is that of an axion. If only one axion is added, Majorana neutrinos still lead to undesired AdS vacua and
would be ruled out. In the case of Dirac neutrinos absence of dangerous vacua are obtained if  the lightest neutrinos have $m_{\nu_1}\leq 4\times 10^{-3}$ eV for NI
($m_{\nu_3}\leq 1\times 10^{-3}$ eV for IH). In this latter case however the axion must have $m_a\geq 2\times 10^{-2}$ eV, so that it cannot be identified
with a standard QCD axion.

The existence of 3D,2D SM vacua can leave an imprint in cosmology (see \cite{Graham:2010hh}). Indeed if our universe came from a lower dimensional
one in 2+1 dimensions there could be some detectable imprints, due to the anisotropy of space.  This may affect the CMB if the last period of
inflation was not too long. This effect would appear at the highest multipoles. However we have just seen that AdS SM vacua cannot be stable, so that 
such anisotropies could not originate from such primordial vacua. Only dS 3D vacua would still be possible, but we have seen that such vacua only
appears for very narrow regions for the mass of the lightest neutrino. Thus e.g. in the case of Dirac neutrinos 3D dS vacua only appear in the
region $6.7\times 10^{-3}eV\leq m_{\nu_1} \leq 7.7\times 10^{-3}eV$ for NI or 
$2.1\times 10^{-3}eV\leq m_{\nu_3} \leq 2.56 \times 10^{-3}eV$ (see Table 1).
 
A further quite important result is that the upper bound on the neutrino masses can be translated into un \emph{upper bound on the EW scale} for fixed cosmological constant and Yukawa couplings. This is a consequence of the dependence of the neutrino masses on the Higgs vev. In  the case of massive Majorana neutrinos with a see-saw mechanism associated to a large scale  $M\simeq 10^{10-14}$ GeV and $Y_{\nu_1}\simeq 10^{-3}$, one obtains that the EW scale cannot exceed $M_{EW}\lesssim 10^2-10^4$ GeV. Similar constraints apply to the Dirac case.  These results are displayed in fig.\ref{fig:vevcc}. From this  perspective, the {\it Higgs scale is small compared to the UV scale because of the smallness of
the c.c. } Parameters yielding higher EW scales would yield lower dimensional AdS vacua and would be inconsistent with an embedding into  quantum gravity. This can bring a new perspective into the issue of the EW hierarchy.  If indeed this is the explanation for the EW hierarchy problem, saturation of the bounds from the WGC provides specific 
predictions for the lightest neutrino mass which are summarized in table \ref{tablapred}. Thus e.g., if a  Majorana mass for the electron neutrino is eventually measured,  values 
$m_{\nu_1}=5(1)\times 10^{-3}$ eV  for NH(IH) would give  a strong indication that the 
present WGC arguments play an important role in the understanding of the EW hierarchy 
problem.  On the other hand the above WGC arguments not necessarily imply the absence of any new physics above the EW scale. Thus e.g. SUSY may be present for 
other reasons like dark matter and in particular the stability of the Higgs potential at higher energies.

We find quite remarkable that a very abstract condition like the absence of stable  AdS vacua may give rise to  such a wealth of implications on
magnitudes of direct physical relevance like the cosmological constant, neutrino masses and even the origin of the EW hierarchy.
In overall,  our analysis is a clear example of how consistency with quantum gravity can have important implications on IR physics. 
Not all points in the parameter space leading to different quantum field theories are allowed when including gravity, and apparent \emph{fine-tuning} problems can turn out to be only mirages due to our ignorance of the actual landscape of consistent theories. This can force us to review our
 notions  of \emph{naturalness} and the hierarchy problems in particle physics when combined with quantum gravity.

\vspace{1.5cm}

\centerline{\bf \large Acknowledgments}

\vspace{0.5cm}

\noindent We thank H. Ooguri for correspondence and
  F. Marchesano, M. Montero,  and A. Uranga for useful discussions. 
The work of L.E.I   has been supported by the ERC Advanced Grant SPLE under contract ERC-2012-ADG-20120216-320421, 
and the  spanish AEI and  the EU FEDER through the projects FPA2015-65480-P and FPA2016-78645-P, 
  as well as  the grant SEV-2012-0249 of the ``Centro de Excelencia Severo Ochoa" Programme.  V.M.L 
  acknowledges support of the Consolider MULTIDARK project CSD2009-00064, the SPLE ERC project  and the
  BMBF under project 05H15PDCAA. I.V. is suported by a grant of the Max Planck society.

\newpage

\appendix


\newpage



\begin{thebibliography}{99}
  



\bibitem{BT}
  J.~D.~Brown and C.~Teitelboim,
  ``Neutralization of the Cosmological Constant by Membrane Creation,''
  Nucl.\ Phys.\ B {\bf 297} (1988) 787;
  ``Dynamical Neutralization of the Cosmological Constant,''
  Phys.\ Lett.\ B {\bf 195} (1987) 177.
  
  
  \bibitem{BP}
  R.~Bousso and J.~Polchinski,
  ``Quantization of four form fluxes and dynamical neutralization of the cosmological constant'',
  JHEP {\bf 0006}, 006 (2000)
  [hep-th/0004134]\\
    J.~L.~Feng, J.~March-Russell, S.~Sethi and F.~Wilczek,
  ``Saltatory relaxation of the cosmological constant,''
  Nucl.\ Phys.\ B {\bf 602} (2001) 307
  [hep-th/0005276].




\bibitem{Weinberg:1988cp} 
  S.~Weinberg,
  ``The Cosmological Constant Problem,''
  Rev.\ Mod.\ Phys.\  {\bf 61}, 1 (1989).
 

\bibitem{ArkaniHamed:2007gg}
  N.~Arkani-Hamed, S.~Dubovsky, A.~Nicolis and G.~Villadoro,
  ``Quantum Horizons of the Standard Model Landscape,''
  JHEP {\bf 0706} (2007) 078
  [hep-th/0703067 [HEP-TH]].


\bibitem{Fornal:2011tw}
  B.~Fornal and M.~B.~Wise,
  ``Standard model with compactified spatial dimensions,''
  JHEP {\bf 1107} (2011) 086
  [arXiv:1106.0890 [hep-th]].


\bibitem{Arnold:2010qz}
  J.~M.~Arnold, B.~Fornal and M.~B.~Wise,
  ``Standard Model Vacua for Two-dimensional Compactifications,''
  JHEP {\bf 1012} (2010) 083
  [arXiv:1010.4302 [hep-th]].


\bibitem{WGC}
  C.~Vafa, ``The String landscape and the swampland,''
  hep-th/0509212\\
  N.~Arkani-Hamed, L.~Motl, A.~Nicolis and C.~Vafa,
  ``The String landscape, black holes and gravity as the weakest force,''
  JHEP {\bf 0706} (2007) 060
  [hep-th/0601001]\\
  H.~Ooguri and C.~Vafa,
  ``On the Geometry of the String Landscape and the Swampland,''
  Nucl.\ Phys.\ B {\bf 766}, 21 (2007)
  [hep-th/0605264].



\bibitem{WGC1}
  T.~Rudelius,
  ``Constraints on Axion Inflation from the Weak Gravity Conjecture,''
  JCAP {\bf 1509} (2015) no.09,  020
  [arXiv:1503.00795 [hep-th]]\\
  M.~Montero, A.~M.~Uranga and I.~Valenzuela,
  ``Transplanckian axions!?,''
  JHEP {\bf 1508} (2015) 032
  [arXiv:1503.03886 [hep-th]]\\
  J.~Brown, W.~Cottrell, G.~Shiu and P.~Soler,
  ``Fencing in the Swampland: Quantum Gravity Constraints on Large Field Inflation,''
  JHEP {\bf 1510} (2015) 023
  [arXiv:1503.04783 [hep-th]]\\
  J.~Brown, W.~Cottrell, G.~Shiu and P.~Soler,
  ``On Axionic Field Ranges, Loopholes and the Weak Gravity Conjecture,''
  JHEP {\bf 1604}, 017 (2016)
  [arXiv:1504.00659 [hep-th]]\\
  B.~Heidenreich, M.~Reece and T.~Rudelius,
  ``Weak Gravity Strongly Constrains Large-Field Axion Inflation,''
  JHEP {\bf 1512} (2015) 108
   [arXiv:1506.03447 [hep-th]]\\
  C.~Cheung and G.~N.~Remmen,
  ``Naturalness and the Weak Gravity Conjecture,''
  Phys.\ Rev.\ Lett.\  {\bf 113} (2014) 051601
   [arXiv:1402.2287 [hep-ph]]\\
  A.~de la Fuente, P.~Saraswat and R.~Sundrum,
  ``Natural Inflation and Quantum Gravity,''
  Phys.\ Rev.\ Lett.\  {\bf 114} (2015) no.15,  151303
    [arXiv:1412.3457 [hep-th]]\\
  A.~Hebecker, P.~Mangat, F.~Rompineve and L.~T.~Witkowski,
  ``Winding out of the Swamp: Evading the Weak Gravity Conjecture with F-term Winding Inflation?,''
  Phys.\ Lett.\ B {\bf 748} (2015) 455
  [arXiv:1503.07912 [hep-th]]\\
  T.~C.~Bachlechner, C.~Long and L.~McAllister,
  ``Planckian Axions and the Weak Gravity Conjecture,''
  JHEP {\bf 1601} (2016) 091
   [arXiv:1503.07853 [hep-th]]\\
  T.~Rudelius,
  ``On the Possibility of Large Axion Moduli Spaces,''
  JCAP {\bf 1504} (2015) no.04,  049
  [arXiv:1409.5793 [hep-th]]\\
  D.~Junghans,
  ``Large-Field Inflation with Multiple Axions and the Weak Gravity Conjecture,''
  JHEP {\bf 1602} (2016) 128
  [arXiv:1504.03566 [hep-th]]\\
  K.~Kooner, S.~Parameswaran and I.~Zavala,
  ``Warping the Weak Gravity Conjecture,''
  Phys.\ Lett.\ B {\bf 759}, 402 (2016)
  [arXiv:1509.07049 [hep-th]]\\
  D.~Harlow,
  ``Wormholes, Emergent Gauge Fields, and the Weak Gravity Conjecture,''
  JHEP {\bf 1601}, 122 (2016)
    [arXiv:1510.07911 [hep-th]]\\
  A.~Hebecker, F.~Rompineve and A.~Westphal,
  ``Axion Monodromy and the Weak Gravity Conjecture,''
  JHEP {\bf 1604} (2016) 157
   [arXiv:1512.03768 [hep-th]].




\bibitem{WGC2} 
  M.~Montero, G.~Shiu and P.~Soler,
  ``The Weak Gravity Conjecture in three dimensions,''
  arXiv:1606.08438 [hep-th]\\
    B.~Freivogel and M.~Kleban,
  ``Vacua Morghulis,''
  arXiv:1610.04564 [hep-th]\\
  P.~Saraswat,
  ``The Weak Gravity Conjecture and Effective Field Theory,''
  arXiv:1608.06951 [hep-th]\\
   D.~Klaewer and E.~Palti,
  ``Super-Planckian Spatial Field Variations and Quantum Gravity,''
  arXiv:1610.00010 [hep-th]\\
    L.~McAllister, P.~Schwaller, G.~Servant, J.~Stout and A.~Westphal,
  ``Runaway Relaxion Monodromy,''
  arXiv:1610.05320 [hep-th].



\bibitem{WGC3}
  M.~Montero, A.~M.~Uranga and I.~Valenzuela,
  ``A Chern-Simons Pandemic,''
  arXiv:1702.06147 [hep-th]\\
   W.~Cottrell, G.~Shiu and P.~Soler,
  ``Weak Gravity Conjecture and Extremal Black Holes,''
  arXiv:1611.06270 [hep-th]\\
    A.~Hebecker, P.~Henkenjohann and L.~T.~Witkowski,
  ``What is the Magnetic Weak Gravity Conjecture for Axions?,''
  Fortsch.\ Phys.\  {\bf 65} (2017) no.3-4,  1700011
  doi:10.1002/prop.201700011
  [arXiv:1701.06553 [hep-th]]\\
    A.~Hebecker and P.~Soler,
  ``The Weak Gravity Conjecture and the Axionic Black Hole Paradox,''
  arXiv:1702.06130 [hep-th]\\
  E.~Palti,
  ``The Weak Gravity Conjecture and Scalar Fields,''
  arXiv:1705.04328 [hep-th].
 



\bibitem{OV}
H.~Ooguri and C.~Vafa,
``Non-supersymmetric AdS and the Swampland,''
arXiv:1610.01533 [hep-th].


\bibitem{MoreOV}
  U.~Danielsson and G.~Dibitetto,
  ``The fate of stringy AdS vacua and the WGC,''
  arXiv:1611.01395 [hep-th]\\
    T.~Banks,
  ``Note on a Paper by Ooguri and Vafa,''
  arXiv:1611.08953 [hep-th]\\
    H.~Ooguri and L.~Spodyneiko,
  ``New Kaluza-Klein Instantons and Decay of AdS Vacua,''
  arXiv:1703.03105 [hep-th].
  

\bibitem{Maldacena:1998uz}
  J.~M.~Maldacena, J.~Michelson and A.~Strominger,
  ``Anti-de Sitter fragmentation,''
  JHEP {\bf 9902} (1999) 011
  [hep-th/9812073].

\bibitem{Barbon:2010gn}
  J.~L.~F.~Barbon and E.~Rabinovici,
  ``Holography of AdS vacuum bubbles,''
  JHEP {\bf 1004} (2010) 123
  doi:10.1007/JHEP04(2010)123
  [arXiv:1003.4966 [hep-th]].
\bibitem{Harlow:2010az}
  D.~Harlow,
  ``Metastability in Anti de Sitter Space,''
  arXiv:1003.5909 [hep-th].


\bibitem{coleman}
S.~R. Coleman and F.~De~Luccia, {\it {Gravitational Effects on and of Vacuum
  Decay}},  {\em Phys. Rev.} {\bf D21} (1980) 3305.

\bibitem{KKLT}
  S.~Kachru, R.~Kallosh, A.~D.~Linde and S.~P.~Trivedi,
  ``De Sitter vacua in string theory,''
  Phys.\ Rev.\ D {\bf 68} (2003) 046005
  [hep-th/0301240].

\bibitem{deAlwis:2013gka}
  S.~de Alwis, R.~Gupta, E.~Hatefi and F.~Quevedo,
  ``Stability, Tunneling and Flux Changing de Sitter Transitions in the Large Volume String Scenario,''
  JHEP {\bf 1311} (2013) 179
  [arXiv:1308.1222 [hep-th]].


\bibitem{Clifton:2007en}
  T.~Clifton, A.~D.~Linde and N.~Sivanandam,
  ``Islands in the landscape,''
  JHEP {\bf 0702} (2007) 024
  [hep-th/0701083].

\bibitem{Brown:2010mg}
  A.~R.~Brown and A.~Dahlen,
  ``Giant Leaps and Minimal Branes in Multi-Dimensional Flux Landscapes,''
  Phys.\ Rev.\ D {\bf 84} (2011) 023513
  [arXiv:1010.5241 [hep-th]].

\bibitem{Brown:2011ry}
  A.~R.~Brown and A.~Dahlen,
  ``Populating the Whole Landscape,''
  Phys.\ Rev.\ Lett.\  {\bf 107} (2011) 171301
  [arXiv:1108.0119 [hep-th]].

 
\bibitem{Witten}
  E.~Witten,
  ``Instability of the Kaluza-Klein Vacuum,''
  Nucl.\ Phys.\ B {\bf 195} (1982) 481.

 
 
\bibitem{Olive:2016xmw}
  C.~Patrignani {\it et al.} [Particle Data Group],
  ``Review of Particle Physics,''
  Chin.\ Phys.\ C {\bf 40} (2016) no.10,  100001.

 
 
\bibitem{Simpson:2017qvj}
  F.~Simpson, R.~Jim\'enez, C.~Pe\~na-Garay and L.~Verde,
  ``Strong Evidence for the Normal Neutrino Hierarchy,''
  arXiv:1703.03425 [astro-ph.CO].

\bibitem{Agostini:2017jim}
  M.~Agostini, G.~Benato and J.~Detwiler,
  ``Discovery probability of next-generation neutrinoless double-$\beta$ decay experiments,''
  arXiv:1705.02996 [hep-ex].

  
\bibitem{Schwetz:2017fey}
  T.~Schwetz, K.~Freese, M.~Gerbino, E.~Giusarma, S.~Hannestad, M.~Lattanzi, O.~Mena and S.~Vagnozzi,
  ``Comment on "Strong Evidence for the Normal Neutrino Hierarchy",''
  arXiv:1703.04585 [astro-ph.CO].
\bibitem{Vagnozzi:2017ovm}
  S.~Vagnozzi, E.~Giusarma, O.~Mena, K.~Freese, M.~Gerbino, S.~Ho and M.~Lattanzi,
  ``Unveiling $\nu$ secrets with cosmological data: neutrino masses and mass hierarchy,''
  arXiv:1701.08172 [astro-ph.CO].

\bibitem{sterile}
  S.~Gariazzo, C.~Giunti, M.~Laveder, Y.~F.~Li and E.~M.~Zavanin,
  ``Light sterile neutrinos,''
  J.\ Phys.\ G {\bf 43} (2016) 033001
  doi:10.1088/0954-3899/43/3/033001
  [arXiv:1507.08204 [hep-ph]]\\
  K.~N.~Abazajian {\it et al.},
  ``Light Sterile Neutrinos: A White Paper,''
  arXiv:1204.5379 [hep-ph].


\bibitem{sterilelight}
  A.~Palazzo,
  ``Constraints on very light sterile neutrinos from $\theta_{13}$-sensitive reactor experiments,''
  JHEP {\bf 1310} (2013) 172
  [arXiv:1308.5880 [hep-ph]].


\bibitem{Oyama:2016lor}
  Y.~Oyama and M.~Kawasaki,
  ``Constraining light gravitino mass with 21 cm line observation,''
  arXiv:1605.09191 [astro-ph.CO].

\bibitem{Osato:2016ixc}
  K.~Osato, T.~Sekiguchi, M.~Shirasaki, A.~Kamada and N.~Yoshida,
  ``Cosmological Constraint on the Light Gravitino Mass from CMB Lensing and Cosmic Shear,''
  JCAP {\bf 1606} (2016) no.06,  004
  [arXiv:1601.07386 [astro-ph.CO]].
  
  
  
\bibitem{Brust:2013xpv}
  C.~Brust, D.~E.~Kaplan and M.~T.~Walters,
  ``New Light Species and the CMB,''
  JHEP {\bf 1312} (2013) 058
  [arXiv:1303.5379 [hep-ph]].


\bibitem{Pospelov:2010hj}
  M.~Pospelov and J.~Pradler,
  ``Big Bang Nucleosynthesis as a Probe of New Physics,''
  Ann.\ Rev.\ Nucl.\ Part.\ Sci.\  {\bf 60} (2010) 539
  [arXiv:1011.1054 [hep-ph]].


  
  
  
  
  

\bibitem{Maltoni:2015twa}
  F.~Maltoni, A.~Martini, K.~Mawatari and B.~Oexl,
  ``Signals of a superlight gravitino at the LHC,''
  JHEP {\bf 1504} (2015) 021
  [arXiv:1502.01637 [hep-ph]].



\bibitem{Ostrovskiy:2016uyx}
  I.~Ostrovskiy and K.~O'Sullivan,
  ``Search for neutrinoless double beta decay,''
  Mod.\ Phys.\ Lett.\ A {\bf 31} (2016) no.18,  1630017
   Erratum: [Mod.\ Phys.\ Lett.\ A {\bf 31} (2016) no.23,  1692004]
  [arXiv:1605.00631 [hep-ex]].


 
 

\bibitem{villadoro}
  G.~Grilli di Cortona, E.~Hardy, J.~Pardo Vega and G.~Villadoro,
  ``The QCD axion, precisely,''
  JHEP {\bf 1601} (2016) 034
  [arXiv:1511.02867 [hep-ph]].


 
 
 
\bibitem{relaxion}
  P.~W.~Graham, D.~E.~Kaplan and S.~Rajendran,
  ``Cosmological Relaxation of the Electroweak Scale,''
  Phys.\ Rev.\ Lett.\  {\bf 115} (2015) no.22,  221801,
   [arXiv:1504.07551 [hep-ph]]\\
   J.~R.~Espinosa, C.~Grojean, G.~Panico, A.~Pomarol, O.~Pujolˆs and G.~Servant,
  ``Cosmological Higgs-Axion Interplay for a Naturally Small Electroweak Scale,''
  Phys.\ Rev.\ Lett.\  {\bf 115} (2015) no.25,  251803
  [arXiv:1506.09217 [hep-ph]].


    
  \bibitem{ourrelaxion}
  L.~E.~Ib\'a\~nez, M.~Montero, A.~Uranga and I.~Valenzuela,
  ``Relaxion Monodromy and the Weak Gravity Conjecture,''
  arXiv:1512.00025 [hep-th]\\
   A.~Hebecker, J.~Moritz, A.~Westphal and L.~T.~Witkowski,
  ``Axion Monodromy Inflation with Warped KK-Modes,''
  arXiv:1512.04463 [hep-th].
    
    
    \bibitem{hierarxion}
  A.~Herr\'aez and L.~E.~Ib\'a\~nez,
  ``An Axion-induced SM/MSSM Higgs Landscape and the Weak Gravity Conjecture,''
  JHEP {\bf 1702} (2017) 109
  [arXiv:1610.08836 [hep-th]].


\bibitem{Hui:2016ltb}
  L.~Hui, J.~P.~Ostriker, S.~Tremaine and E.~Witten,
  ``Ultralight scalars as cosmological dark matter,''
  Phys.\ Rev.\ D {\bf 95} (2017) no.4,  043541
  [arXiv:1610.08297 [astro-ph.CO]].

\bibitem{axiverse}
  A.~Arvanitaki, S.~Dimopoulos, S.~Dubovsky, N.~Kaloper and J.~March-Russell,
  ``String Axiverse,''
  Phys.\ Rev.\ D {\bf 81} (2010) 123530
  [arXiv:0905.4720 [hep-th]].



\bibitem{Graham:2010hh}
  P.~W.~Graham, R.~Harnik and S.~Rajendran,
  ``Observing the Dimensionality of Our Parent Vacuum,''
  Phys.\ Rev.\ D {\bf 82} (2010) 063524
  [arXiv:1003.0236 [hep-th]].



 
\end{thebibliography}
\end{document}